\documentclass[a4paper,fleqn,usenatbib]{mnras}
\usepackage{amsmath,graphicx,amssymb,aas_macros}
\usepackage{txfonts} 
\usepackage{hyperref,nicefrac}

\usepackage{multirow}
\renewcommand{\[}{\begin{equation}}
\renewcommand{\]}{\end{equation}}

\let\boldgrk=\gkvecten
\let\boldgrksc=\gkvecseven

\def\gkthing#1{{\mathchoice%
	{\hbox{{\boldgrk\char#1}}}
	{\hbox{{\boldgrk\char#1}}}
	{\hbox{{\boldgrksc\char#1}}}
	{\hbox{{\boldgrksc\char#1}}}}}

\def\vtheta{\gkthing{18}}

\def\vmu{\gkthing{22}}

{\newif\ifnotend
\notendtrue
\def\veclist{ABCDEFGHIJKLMNOPQRSTUVWXYZabcdefghijklmnopqrstuvwxyz.}
\def\top#1#2.{#1}
\def\tail#1#2.{#2.}
\loop\expandafter\xdef\csname v\expandafter\top\veclist\endcsname%
{{\noexpand\bf\expandafter\top\veclist}}
\edef\veclist{\expandafter\tail\veclist}
\if\veclist.\notendfalse\fi\ifnotend\repeat}
\def\d{{\rm d}}

\def\bolth{\mbox{\boldmath$\theta$}}

\def\Gyr{\,\mathrm{Gyr}}
\def\Myr{\,\mathrm{Myr}}
\def\kpc{\,\mathrm{kpc}}
\def\Mpc{\,\mathrm{Mpc}}

\def\kms{\,\mathrm{km\,s}^{-1}}

\def\mas{\,\mathrm{mas}}
\def\Gevdens{\,\mathrm{GeV\,cm}^{-3}}
\def\msun{\,{\rm M}_\odot}

\def\vsol{\mbox{\boldmath{$v_\odot$}}}
\def\vsfr{\mbox{\boldmath{$v_{\rm SFR}$}}}
\def\alphaFe{[\alpha/\mathrm{Fe}]}
\def\vlos{{v_\parallel}}

\def\rh{r_\mathrm{h}}
\def\pc{\,\mathrm{pc}}
\def\d{\mathrm{d}}

\def\HI{H\,\textsc{i}}
\def\Gaia{{\it Gaia}}

\title[Mass \& gravitational potential of the Milky Way]
{The mass distribution and gravitational potential of the Milky Way}

\author[P.~J.~McMillan]{
  Paul~J.~McMillan\thanks{E-mail: paul@astro.lu.se}\\  
  Lund Observatory, Lund University, Department of Astronomy and Theoretical Physics, Box 43, SE-22100, Lund, Sweden;\\ 
    Rudolf Peierls Centre for Theoretical Physics, 1 Keble Road, Oxford, OX1 3NP, UK
}

\date{Accepted 2016 October 21. Received 2016 October 21; in original form 2016 August 1}

\pubyear{2016}

\begin{document}
\label{firstpage}
\pagerange{\pageref{firstpage}--\pageref{lastpage}}
\maketitle

\begin{abstract}
  We present mass models of the Milky Way created to fit observational
  constraints and to be consistent with expectations from theoretical
  modelling. The method used to create these models is that
  demonstrated in \cite{PJM11:Mass}, and we improve on those models by
  adding gas discs to the potential, considering the effects of
  allowing the inner slope of the halo density profile to vary, and
  including new observations of maser sources in the Milky Way amongst
  the new constraints. We provide a best fitting model, as well as estimates of the properties of the Milky Way. Under the assumptions in our main model, we find that the Sun is $R_0 = (8.20\pm0.09)\kpc$ from the Galactic Centre, with the circular speed at the Sun being $v_0 = (232.8\pm3.0)\kms$; that the Galaxy has a total stellar mass of $(54.3\pm5.7)\times10^9\msun$, a total virial mass of $(1.30 \pm 0.30)\times10^{12}\msun$ and a local dark-matter density of $0.38\pm0.04\Gevdens$, where the quoted uncertainties are statistical. These values are sensitive to our choice of priors and constraints. We investigate systematic uncertainties, which in some cases may be larger. For example, if we weaken our prior on $R_0$, we find it to be $(7.97\pm0.15)\kpc$ and that $v_0=(226.8\pm4.2)\kms$. We find that most of these properties, including the local dark-matter density, are remarkably insensitive to the assumed power-law density slope at the centre of the dark-matter halo. We find that it is unlikely that the local standard of rest differs significantly from that found under assumptions of axisymmetry. We have made code to compute the force from our potential, and to integrate orbits within it, publicly available.
\end{abstract}


\section{Introduction}\label{sec:intro}
The field of Milky Way dynamics is reaching an incredibly exciting time, as the successful launch and operation of the European Space Agency's (ESA's) \Gaia\ satellite \citep{GaiaMissionArXiv} mean that we will soon have access to proper motions and parallaxes for a billion stars. This represents an increase of four orders of magnitude over the number of stars with known parallaxes from \Gaia's predecessor, {\it Hipparcos} \citep{HipparcosCatalogue}. This information will revolutionise how we understand our own Galaxy, and by extension, the Universe as a whole.

However, \Gaia\ can only measure the present velocities of stars, not their accelerations due to the Galactic gravitational potential. These accelerations are far too small, of the order of $\mathrm{cm}\,\mathrm{s}^{-1}\,\mathrm{yr}^{-1}$. This means that the Galaxy's gravitational field can only be inferred, not measured. The stars seen by \Gaia\ orbit in the potential of the Milky Way, and it is the nature of their orbits \citep[best characterised by their actions, e.g.][]{JJBPJM16} that are generally of real interest for understanding Galactic structure, rather than their exact positions and velocities at this moment. To determine the stars' orbits, we need to know the underlying potential.

This paper follows a long tradition of authors who have produced mass models of the Milky Way, with the intention of synthesising all of the knowledge about the components of the Milky Way into a coherent picture of the gravitational potential of the Galaxy. A famous early example was that of \cite{Sc56}, with further examples being provided by \cite{CaOs81},  \citet[henceforth DB98]{WDJJB98:mass}, \cite*{KlZhSo02} and the previous paper in this series \cite[henceforth Paper I]{PJM11:Mass}. 

We return to this subject for three main reasons. Firstly because our knowledge of the Milky Way has increased substantially since Paper I was written, with new constraints that should be synthesised to produce a superior model. Secondly, Paper I did not include a component representing the Milky Way's cold gas. This gas forms a vertically thin component in the Galactic mid-plane, which deepens the potential well close to the plane, significantly affecting the dynamics of stars in the Solar neighbourhood. This means it is important to include this component. Thirdly, \Gaia\ is due to release data in the very near future, which will dramatically increase our knowledge of the positions and velocities of stars in the Milky Way. It is therefore useful to have a model that reflects our current knowledge of the Milky Way's potential, to allow us to calculate best estimates of the orbits of these stars \citep[and their properties, such as the actions][]{SaJJB16}. It also provides a helpful estimate of the potential that can be refined by dynamical modelling \citep[e.g.][]{PJMJJB13,BoRi13,Piea14}.

To find the gravitational potential associated with a given mass model
we use the publicly licensed code \textsc{GalPot}, which is described
by DB98. We have made an edited version of this software available on GitHub, along with files giving the parameters of the best fitting model potentials found in this study in a form that \textsc{GalPot} can read. We also provide routines to integrate orbits in this potential.\footnote{See \url{https://github.com/PaulMcMillan-Astro/GalPot}} In the appendix we give examples of using this code.

In Section~\ref{sec:comp} we describe the components of our model of the Milky Way, and some of the constraints we apply to them. In Section~\ref{sec:data} we describe the kinematic data that our model has to fit, and in Section~\ref{sec:fit} explain how we perform the model fits. Section~\ref{sec:Main} gives the properties of our main model and discusses some of its implications, and in Section~\ref{sec:alternative} we explore alternative models. Finally in Section~\ref{sec:others} we compare our results to those of other authors (and discuss reasons that they may differ), before drawing conclusions in Section~\ref{sec:conc}. 

\section{Components of the Milky Way}\label{sec:comp}
We decompose the Milky Way into 6 axisymmetric components - bulge; dark-matter halo; thin and thick stellar discs; and \HI\ and molecular gas discs. This is similar to that used in Paper I, with the addition of the gas discs. We recap the main properties briefly, and give details of
the components that are new (or have different properties) in this model.

\subsection{The bulge}\label{sec:bulge}

The Galactic bulge has been known for some time to be a
triaxial rotating bar with its long axis in the plane of
the Galaxy \citep*[e.g.][]{JJBGeSp97}. It is increasingly clear that it is a boxy (or `peanut shaped') bulge \citep[and others]{McWZo10,Naea10,Neea12}.  

Our model is axisymmetric for ease of calculation. We must therefore accept that it cannot
accurately represent the inner few kpc of the Galaxy, and we should
not expect it to reproduce measurements taken from that part of the
Galaxy. It is difficult to take constraints on our model from studies of the bulge as they naturally reflect its triaxial shape. Indeed, as noted by \cite{Poea15M2M}, different studies of the bulge constrain the mass in different regions, and use different definitions of what constitutes the bulge.

Our density profile is (as in Paper I) based on the parametric model fit by \cite{BiGe02} to dereddened L-band COBE/DIRBE data
\citep*{SpMaBl96}, and the mass-to-light ratio determined by 
\citeauthor{BiGe02} from a comparison between gas dynamics
in models and those observed in the inner Galaxy. We note that the more recent (non-parametric) study of \cite{Poea15M2M} states that the total bulge mass that they find compares well with that of \cite{BiGe02}, when they consider the region $R<2.2\kpc$. We have chosen not to directly apply the \cite{Poea15M2M} constraint on the total mass in the bulge to our model because it only describes the inner $2.2\kpc$, while our bulge model has around $20$ per cent of its mass outside that radius. We compare our result to the \citeauthor{Poea15M2M} result in Section~\ref{sec:others}, and find reasonable agreement, which is as good as we can expect given the simplifications of our model.

The \citeauthor{BiGe02} model is not axisymmetric, so we make an
axisymmetric approximation which has the density profile
\begin{equation}\label{eq:bulge}
  \rho_\mathrm{b}=\frac{\rho_{0,\mathrm{b}}}{(1+r^\prime/r_0)^\alpha}\;
  \textrm{exp}\left[-\left(r^\prime/r_{\mathrm{cut}}\right)^2\right],
\end{equation}
where, in cylindrical coordinates,
\begin{equation}
  r^\prime = \sqrt{R^2 + (z/q)^2}
\end{equation}
with $\alpha=1.8$, $r_0=0.075\kpc$, $r_{\mathrm{cut}}=2.1\kpc$,
and axis ratio $q=0.5$.  We take a total bulge mass $M_\mathrm{b}=8.9\times 10^9\,\msun$, with an
uncertainty of $\pm10$ per cent. For this density profile, this corresponds to a
scale density of $\rho_{0,\mathrm{b}}=9.93\times 10^{10}\,\msun\kpc^{-3}
\pm 10$ per cent.

\subsection{The stellar discs}\label{sec:stellardisc}

The Milky Way's stellar disc is commonly decomposed into a thin and thick
disc \citep[e.g.][]{GiRe83}. These are modelled as exponential in the
sense that.
\begin{equation}\label{eq:disc}
  \rho_\mathrm{d}(R,z)=\frac{\Sigma_{0}}{2z_\mathrm{d}}\;\textrm{exp}\left(-\frac{\mid
      z\mid}{z_\mathrm{d}}-\frac{R}{R_\mathrm{d}}\right),
\end{equation}
with scale height $z_\mathrm{d}$, scale length $R_\mathrm{d}$ and central surface
density $\Sigma_{0}$. The total disc mass is
$M_\mathrm{d}=2\pi\Sigma_{0}R_\mathrm{d}^2$. We choose not to consider the possibility of a central `hole' in the stellar density for the same reasons given in Paper I.

The \cite{Juea08} analysis of data from the Sloan Digital Sky Survey
\citep[SDSS:][]{SDSS7} showed that the approximation to
exponential profiles is a sensible one for the Milky Way, and produced
estimates based on photometry for the scale lengths, scale heights and
relative densities of the two discs. As in Paper I use these values as
constraints. We hold the scale heights 
of the discs fixed as $z_{\mathrm{d},\mathrm{thin}}=300\pc$ and 
$z_{\mathrm{d},\mathrm{thick}}=900\pc$. In Paper I we showed that
our models are not significantly affected by the choice of
scale heights. We take the scale lengths for the thin and thick
discs to be $(2.6\pm0.52)\kpc$ and $(3.6\pm0.72)\kpc$ and the local density normalisation $f_{\mathrm{d},\odot} =
\rho_{\mathrm{thick}}(R_\odot,z_\odot)/ \rho_{\mathrm{thin}}(R_\odot,z_\odot)$ is 
taken to be $0.12\pm0.012$. For a discussion of the many available studies of these parameters, see the review of the properties of the Milky Way by \cite{BHGe16}.

Recent studies such as those of \cite{Beea11},  \cite{Boea12_MAPdensity} and \cite{Anea14} have shown that when the `thick  disc' of the Galaxy is defined chemically (as comprising stars with high $\alphaFe$), it clearly has a shorter scale length than the `thin disc' (comprised of stars with low $\alphaFe$). Since the high $\alphaFe$ component also has a larger scale height than the low $\alphaFe$ population, this might appear to be in conflict with the \cite{Juea08} result, and with thick discs observed in external galaxies \citep[e.g.][]{YoDa06}, where the thick disc has the longer scale length. However, this is simply because there is a distinction between defining the thick disc chemically, and defining it morphologically, easily explained if the chemically-defined discs are flared \citep[e.g.][]{Miea15}. Since we are only interested here in the morphology of the discs (and therefore their potential), we can happily accept the \citeauthor{Juea08} result and not concern ourselves with the chemical properties of the two components we label as separate stellar discs.

\subsection{The gas discs}\label{sec:gasdisc}

The models used in Paper I contained no component representing gas in
the Milky Way. Simple dynamical arguments demonstrate that this is a mistake when the model is used to understand stellar dynamics in the Solar neighbourhood. Since the gas discs have a
much smaller scale height than the stellar component near the Sun,
its presence significantly deepens the potential well near the Sun,
even if the total surface density remains unchanged. This deeper well
means that stars that reach far from the Galactic plane have large
velocities in the $z$ direction when they pass near the Sun. The
potential due to the gas disc is therefore necessary to make the
number of stars with high $v_z$ in the Solar neighbourhood dynamically
consistent with the observed density of stars far from the Galactic
plane.

In this study we therefore improve upon Paper I by including two components
representing the \HI\ and molecular gas discs of the Milky Way. These
discs follow the density law\footnote{This is the same form as was used by DB98, but the expression given in the
published journal version of that paper contains a typographical
error (the term $R_\mathrm{h}/R$ is mistakenly given as 
$R_\mathrm{h}/R_\mathrm{d}$). This error does not appear in the arXiv
version of the article: astro-ph/9612059.} 
\begin{equation}\label{eq:gasdisc}
  \rho_\mathrm{d}(R,z)=\frac{\Sigma_{0}}{4z_\mathrm{d}}\;
  \textrm{exp}\left(-\frac{R_{\rm m}}{R}-\frac{R}{R_\mathrm{d}}\right)\;
  {\rm sech}^2(z/2z_\d),
\end{equation}
which, like the stellar disc model (equation~\ref{eq:disc}), declines exponentially with $R$ at large
$R$, but also has a hole in the centre with associated scale length
$R_{\rm m}$. The maximum surface density is found at $R = \sqrt{R_\d
  R_{\rm m}}$, and the total disc mass is $M_\d = 2\pi \Sigma_{0} R_\d
R_{\rm m} K_2(2\sqrt{R_{\rm m}/R_\d})$ where $K_2$ is a modified
Bessel function. The disc model has an `isothermal' ${\rm sech}^2$
vertical profile. Note that, following DB98, we include a
  factor of $1\over2$ in the ${\rm sech}^2$ term. This ensures that as
  $z\rightarrow\infty$ the density goes as $\exp -z/z_\d$.

There is still a great deal of uncertainty over the large scale distribution of
gas in the Milky Way (see discussions by \citealt{Lo02} and \citealt{KaDe08}, henceforth KD08).\footnote{For example, \cite{Lo02} note that the surface density plot from \cite{Da93}, which has been reproduced without comment in standard texts, includes values that have been arbitrarily scaled by a factor of 2.} Both
the \HI\ and molecular gas discs have `holes'' with little gas in the
inner few kpc, and the \HI\ disc is significantly more extended in
$R$ and has a greater scale height. 

Our \HI\ disc model is designed to resemble the distribution found by
KD08, but with significant simplifications,
made in order to greatly simplify the force calculation using
\textsc{GalPot}. The largest simplification is that we neglect the
flaring of the gas disc, and instead keep a constant scale height
$z_{\rm HI}=85\pc$, consistent with the half-width half-maximum
distance of $150\pc$ at $R_0$ given by KD08. We set the surface
density to be $10 \msun\pc^{-2}$ at a fiducial value of
$R_0=8.33\kpc$, consistent with the value given by KD08. It is worth noting that the KD08 value is the total surface density of \emph{all} the gas associated with \HI, i.e. not just the hydrogen (Kalberla priv. comm.). This is not made clear in KD08, and this has led some authors to (mistakenly) add a ${\sim}40$ per cent correction for helium and metals \citep[e.g.][]{He15}. We
set $R_{\rm m,HI}=4\kpc$ and $R_{\rm d,HI}=7\kpc$, which produces a
surface density that varies more smoothly than that of KD08 (which
has a constant surface density for $4\kpc\lesssim R\lesssim12.5\kpc$,
and an exponential decline for $R\gtrsim12.5$), but is broadly
similar, and shares the property of having ${\sim}21$ per cent of the \HI\
mass at $R<R_0$. The total mass in the \HI\ component (including helium
and metals) is $1.1\times10^{10}\msun$.

Our molecular gas disc is intended to resemble that described by
\cite{Da93} and \cite{OlMe01} in terms of scale height and surface density at the Solar position, and the Galactocentric radius of the peak in the surface density. We give it a constant scale height
$z_{\rm H_2}=45\pc$, consistent with the full-width half-maximum value
of ${\sim}160\pc$ at $R_0$ given by \citeauthor{OlMe01}. The surface
density at the fiducial radius $R_0=8.33\kpc$ is set as
$2\msun\pc^{-2}$, with scale length $R_{\rm d,H_2}=1.5\kpc$ and
$R_{\rm m,H_2} = 12\kpc$, which places the maximum surface density at
$R=4\kpc$. The total mass in the molecular gas component is $1.2\times10^{9}\msun$.

\begin{table}
  \begin{tabular}{ccccccccc}
    \hline
    \multirow{2}{*}{Disc} & $R_\d$ & $R_{\rm m}$ & $z_\d$ & & $\Sigma_0$ & $\Sigma_\odot$ &
    & $M$ \\
    &\multicolumn{3}{c}{($\kpc$)} & &
    \multicolumn{2}{c}{$(\msun\pc^{-2})$} & & $(\msun)$ \\ 
    \hline
    \HI\ & $7$ & $4$ & $0.085$ &  & $53.1$ & $10$ & & $1.1\times10^{10}$ \\
    H$_2$ & $1.5$ & $12$ & $0.045$ &  & $2180$ & $2$ & & $1.2\times10^9$ \\
  \end{tabular}
  \caption{
    Parameters of the gas discs in the mass models. The first four
    columns enter the description of the disc in eq.~\ref{eq:gasdisc},
    from which one can derive the values in the final two columns.
  }
  \label{tab:gas}
\end{table}

\begin{figure}
  \centerline{\resizebox{\hsize}{!}{\includegraphics[angle=270]{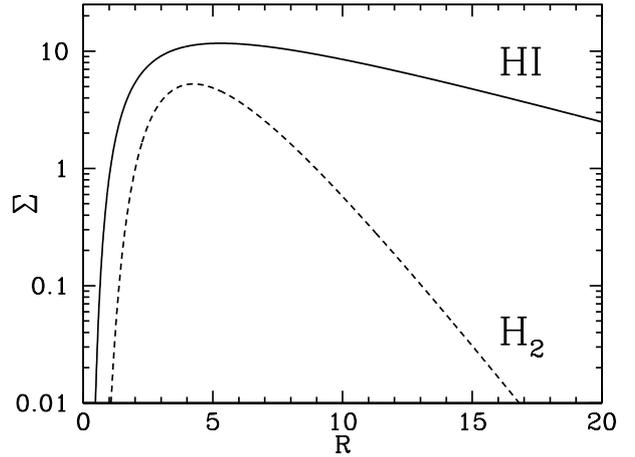}}}
  \caption{ Surface densities, as a function of radius, of the two gas discs used in the model.
    \label{fig:gas}
  }
\end{figure}

The parameters of these two discs are given in Table~\ref{tab:gas}. We
show the corresponding surface density profiles of these two discs in
Fig.~\ref{fig:gas}. 

\subsection{The dark-matter halo} \label{sec:halo} 

We describe our halo density with the simple density profile
\begin{equation} \label{eq:halo} \rho_{\mathrm{h}} =
  \frac{\rho_{0,\mathrm{h}}}{x^\gamma\,(1+x)^{3-\gamma}},
\end{equation}
where $x=r/r_{\mathrm{h}}$, with $r_{\rm h}$ the scale radius. For
our main model we follow Paper I and take $\gamma=1$, which is the NFW
profile \citep*{NFW96} that is a common approximation to the density profile found in dark-matter-only
cosmological simulations. 

It is still unclear what effect the baryons in galaxies have on
dark-matter profiles. The main influence will be at the centre, but
whether, for Milky Way sized galaxies, this influence will be to
steepen the inner density `cusp' (corresponding to $\gamma>1$ in
eq.~\ref{eq:halo}) or to weaken it, producing more of a `core'
($\gamma<1$) is an active area of research
\citep[e.g.][]{Duea10,Goea12}. In Section~\ref{sec:vargamma} we explore the
effect of varying the assumed value of $\gamma$.

Haloes in dark-matter-only cosmological simulations tend to be
significantly prolate, but with a great deal of variation in axis
ratios \citep[e.g.][]{Alea06}. It is recognised that, again, baryonic
physics will play an important role -- condensation of baryons to the
centres of haloes is expected to make them rounder than dark-matter-only 
simulations would suggest \citep{Deea08}.  The shape of the Milky
Way's halo is still very much the subject of debate, with different
efforts to fit models of the Sagittarius dwarf's orbit favouring
conflicting halo shapes \citep*[see e.g.][]{LaMaJo09}. In this study we will only consider spherically symmetric haloes.


We take as a prior the relationship between the total stellar mass
$M_\ast$ (the sum of the mass of the bulge and stellar discs) and the virial mass $M_{v}$ found by \cite*{Moea13}. The virial mass is the mass within the virial
radius $r_{\rm v}$, which in this case is defined as the radius of a sphere,
centred on the centre of the galaxy, that has an average density $200$
times the critical density. The critical density is
\begin{equation}
\rho_{\rm crit} = \frac{3\mathrm{H}^2}{8\pi G}
\end{equation}
where we take $H=70.4\kms\Mpc^{-1}$ from \cite{WMAP7}.
The relationship between $M_\ast$ and $M_{v}$ was determined by `abundance matching' of the galaxy stellar masses found by \cite{LiWh09} with the halo virial masses found in the Millennium simulations \citep{Mill05,MillII09}. The relationship is given by 
\begin{equation} \label{eq:mast} 
M_\ast = M_{v} \times
  2\,N\left[\left(\frac{M_{v}}{M_1}\right)^{-\beta} +
    \left(\frac{M_{v}}{M_1}\right)^{\gamma}\right]^{-1},
\end{equation}
with best fitting parameters at redshift $z=0$ of $N=0.0351$, $\log_{10} M_1=11.59$, $\beta=1.376$ and $\gamma=0.608$. 

\cite{Moea13} assume an intrinsic scatter around this relationship of $0.15$ in $\log_{10}M_\ast$, and also give a `plausible range' for all of the parameters in equation~\ref{eq:mast}. It is not straightforward to convert a range of plausible parameters into an uncertainty in the typical stellar mass at a given $M_v$ (which is what we require), especially as we do not know the correlations between the various parameters (e.g. are certain values of $N$ only plausible for certain other values of $M_1$?). Based on the ranges given we can conservatively estimate that the uncertainty in the best fitting value is ${\sim}0.15$ in $\log_{10}M_\ast$  (i.e. similar to the intrinsic spread). Combining this uncertainty of the best fitting value in quadrature with the expected scatter around the best fitting value, we come to a total uncertainty in $\log_{10}M_\ast$ of $\pm0.2$ about the value given by equation~\ref{eq:mast}

Once the shape of a dark-matter halo profile has been chosen (e.g. the
form of eq.~\ref{eq:halo} and a value for $\gamma$), there are still
typically two parameters required to describe it. In the case of
eq.~\ref{eq:halo} these are the scale radius and density, but it is
common for cosmologists to use instead the virial mass and the concentration
$c_{\rm v}$. The concentration is given by $c_{\rm v}=r_{v}/r_{-2}$,
where $r_{-2}$ is the radius at which the derivative
$\d \log\rho/\d\log r = -2$. In the case of the density profile in
eq~\ref{eq:halo}, $r_{-2} = (2-\gamma)\;r_{\rm h}$.

We take a prior on concentration, $c_{\rm v^\prime}$, from \cite{BKea10}\footnote{There are varying definitions of the
  virial radius, and in this case the definition is the radius of a
  sphere with average density of approximately $94$ times the critical
  density. We refer to this as $c_{\rm v^\prime}$ rather than $c_{\rm
    v}$.}  who found that, for the mass range that the Milky Way's
halo is very likely to lie in, the probability of a given
concentration is Gaussian in $\ln c_{\rm v^\prime}$
\begin{equation}\label{eq:conc}
  \ln c_{\rm v^\prime} = 2.56 \pm 0.272.
\end{equation}
This is the same prior taken in Paper I. It corresponds to a
concentration of $12.9$ with uncertainty of around $30$ per cent.

Baryonic physics is likely to have an effect on the concentration,
much as it does on the inner density profile. \cite{Duea10}, who found
increases in the inner density slope of dark-matter profiles from
baryonic processes, found that the concentration of the haloes were also
altered. Some had higher concentrations and some lower, depending on
the type of cooling or feedback used in the simulations. These changes
were typically ${\sim}20$ per cent. Since we therefore have no clear
indication of what changes in $c_{\rm v}$ we should expect as $\gamma$
changes, and since the uncertainty in our constraint on $c_{\rm v}$ is
larger than the typical changes found by \cite{Duea10}, we take
eq.~\ref{eq:conc} as our prior on $c_{\rm v}$ even when we consider
models with $\gamma\neq1$.

\subsection{The Sun}

There is still some uncertainty about the distance from the
Sun to the Galactic Centre $R_0$ \cite[for a review see][]{BHGe16}. Since the interpretation of all of
the observational data considered in Section~\ref{sec:data} depends on
$R_0$, we leave it as a free parameter in our models.

We take a constraint on $R_0$ from \cite{Chea15}. This combines their own analysis of the dynamics of the Milky Way's nuclear stellar cluster with the study of \cite{Giea09} on stellar orbits around the
supermassive black hole at the Galactic Centre, Sgr A*. It arrives at a combined estimate
\begin{equation} \label{eq:R0} 
R_0 = (8.33\pm0.11)\kpc .
\end{equation}

\section{Kinematic data} \label{sec:data}

\subsection{Maser observations} \label{sec:maser} 

A small but increasing number of Galactic maser sources have been
targeted for extremely accurate astrometric measurements, with
uncertainties of ${\sim}10\,\mu\mathrm{as}$, using very long baseline
interferometry. This allows us to determine the full 6 dimensional
phase space coordinates of these sources to high accuracy, as it
provides parallaxes and proper motions, as well as line-of-sight
velocities. Maser sources are associated with high-mass star forming
regions (HMSFR) which are expected to be on near circular orbits. They
have therefore been used by numerous authors to constrain the
properties of the Milky Way \citep[e.g.,][Paper
I]{Reea09,PJMJJB10:Masers,BoHoRi09,BoBa13,Reea14}. 

\cite{Reea14} summarised the work of a number of groups, most
notably the Bar and Spiral Structure Legacy survey
\citep[BeSSeL,][]{BeSSeL} and VLBI Exploration of Radio Astronomy
\citep[VERA,][]{VERA}, which have determined the parallaxes, proper
motions and line-of-sight velocities of 103 Galactic HMSFRs. This
represents an increase by more than a factor of 4 over the number of
HMSFRs with known parallaxes used in Paper I. \cite{Reea14} used
these data to determine the Sun's position and velocity with a
$\chi^2$ analysis which used the approximation that the HMSFRs were
exactly at the position corresponding to the quoted parallax,
neglecting the uncertainty. We use these data
to constrain our mass model using a version of the likelihood analysis
used by \cite{PJMJJB10:Masers}, and Paper I, which means that 
we calculate the likelihood for each HMSFR given our model as an
integral over the parallax ($\varpi$)
\[
\begin{array}{lll}
\mathcal{L}_{\rm HMSFR} & = & \int \d \varpi\, P(\varpi|{\rm data}) \\
 & & \;\;\int \d^2\vmu\,\d\vlos P(\vmu,\vlos|{\rm data})\;P(\vmu,\vlos|\varpi,{\rm
  Model}) ,
\end{array}
\]
where we have assumed that the quoted errors on $\varpi$, $\vmu$ and
$\vlos$ are uncorrelated -- we also assume that they are Gaussian.
The probability distribution function $P(\vmu,\vlos|\varpi,{\rm
  Model})$ can be directly translated from a probability density
function (pdf) for $\vv_{\rm HMSFR}$ as a function of position in the
Galaxy ($\vx$), given assumed values for $R_0$, $v_0$ (the circular
velocity at $R_0$) and $\vsol$ (the peculiar motion of the Sun with
respect to a circular orbit at $R_0$).

In Paper I we assumed that $P(\vv_{\rm HMSFR}|\vx)$ was a Gaussian pdf
of width $7\kms$ in all directions, centred on the velocity of
circular orbit at $\vx$. In this study we use a more general form for
this pdf. We allow for a typical peculiar velocity for the HMSFRs
$\vsfr$ in Galactocentric cylindrical coordinates such that the pdf
is centred at a velocity
\begin{equation}\label{eq:vbar}
  \overline{\vv}= v_c(R)\mathbf{e_\phi}+\vsfr.
\end{equation}
This is to allow for the possibility that the HMSFRs tend to lag
circular rotation, as claimed in previous studies
\citep{Reea09,BoHoRi09,Reea14}.\footnote{In some of these studies, a
  large fraction of the claimed lag can be explained by the use the
  \cite{WDJJB98:vsol} value for $\vsol$ which is now known to be based
  on false assumptions \citep{SBD10} -- see \cite{PJMJJB10:Masers}.} All three
components of $\vsfr$ are left as free parameters. We also allow for
the possibility that a few of these objects may be strongly forced
from a circular orbit by supernovae \citep[e.g.][]{Saea08}, by
having a two-Gaussian pdf around this central value, with the broader
Gaussian component being there to deal with these outliers.

If we write a three-dimensional Gaussian distribution in $\vv$, centred at
$\overline{\vv}$ and of width (in all directions) $\Delta$ as
$G(\vv,\overline{\vv},\Delta)$, then the pdf we use is
\begin{equation}\label{eq:veldist}
  \begin{array}{lll}
  P(\vv_{\rm HMSFR})\d^3\vv & = & (1-f_{\rm
    out})\times G(\vv_{\rm HMSFR},\overline{\vv},\Delta_{\rm v}) \\
  & & \;+ \;f_{\rm out}
  \times G(\vv_{\rm HMSFR},\overline{\vv},\Delta_{\rm out}), \\
\end{array}
\end{equation}
where $f_{\rm out}$ is an `outlier fraction'. We set $f_{\rm
  out}=0.02$ and $\Delta_{\rm out}=30\kms$. Our results are almost
insensitive to plausible changes in these values. The value of
$\Delta_{\rm v}$ is left as a free parameter. We take a uniform prior in
velocity on all 3 components of $\vsfr$ and on $\Delta_v$.

In order to avoid using any data from HMSFRs likely to be strongly
affected by the Galactic bar, we remove any sources that are likely to
be at $R<R_0/2$ for a fiducial value of $R_0=8.33\kpc$. We perform
this cut by insisting that there is no point within the $1-\sigma$
error bars on parallax for a given object that would place it within
this radial range. This reduces the number of HMSFRs used in the
analysis to $93$. We note that \cite{ChReSo15} emphasise that the assumption of a circular rotation curve is likely to provide false results for the Galactic gravitational potential in the inner ${\sim}4\kpc$, because of the influence of the bar. 

\subsection{Other kinematic data}

The other kinematic data we consider is the same as was used in Paper
I. We summarise it here.

\subsubsection{The Solar velocity}
\label{sec:sun} 
In Paper I we assumed that $\vsol$, the peculiar velocity of the Sun with
respect to a circular orbit at $R_0$, took the value found by \cite*{SBD10}
\begin{equation}\label{eq:vsol}
  \begin{array}{lll}
    \vsol & = &  (U_\odot,V_\odot,W_\odot) \\
    & = & (11.1,12.24,7.25)\kms ,
  \end{array}
\end{equation}
where $U_\odot$ is the velocity towards the Galactic Centre, $V_\odot$
is the velocity in the direction of Galactic rotation and $W_\odot$ is
the velocity perpendicular to the Galactic plane. In this study we
allow $\vsol$ to vary, and treat the \citeauthor*{SBD10} value as a
prior with uncertainty $(1.5,2.0,1.0)\kms$ (taking into account both the
systematic and statistical quoted errors).

To constrain the total velocity of the Sun around the Galactic centre
-- which is the sum of $v_0$, the circular velocity at $R_0$, and
$\vsol$ -- we use the proper motion of Sgr A* in the plane of the
Galaxy, as determined by \cite{ReBr04}
\begin{equation} \label{eq:sgra} 
  \begin{array}{lll}
    \mu_{\mathrm{Sgr A^\ast}} & = & (-6.379\pm 0.026)\,
    \mathrm{mas}\,\mathrm{yr}^{-1} \\
    & =  & (-30.24\pm 0.12)\kms\kpc^{-1} .\\
  \end{array}
\end{equation}
Since Sgr A* is believed to be fixed at the Galactic Centre to within
${\sim}1\kms$, this proper motion is thought to be almost entirely due
to the motion of the Sun around the Galactic Centre,
$(v_0+V_\odot)/R_0$. We add an uncertainty of $2\kms/R_0$ in quadrature
with the observational error to account for uncertainty in the true
velocity of Sgr A*. Note that $v_0$ is not a parameter of the model,
but is defined by the potential.

\subsubsection{Terminal velocity curves} \label{sec:vt} 

As in Paper I we use the terminal velocity curve determined by
\cite{Ma94,Ma95} to constrain the rotation curve for $R<R_0$. We do
this under the assumption that for $|\sin l|>0.5$ (where we assume
that the influence of the bar is negligible) the terminal velocity at
a given Galactic longitude $l$ can be associated with gas on circular
orbits at Galactocentric radius $R = R_0\sin{l}$, so\footnote{Note
  that Paper I's eq. 10 gave an expression for
  $v_{\mathrm{term}}(l)$ that was incorrect both because it did not treat
  $\sin{l}$ correctly in the first term, and because it did not
  explicitly include the correction for $\vsol$ in the expression.}
\begin{equation}
  \label{eq:vterm}
  \begin{array}{lll}
    v_{\mathrm{term}}(l) & = & {\rm
      sgn}(\sin{l})\,v_\mathrm{c}(R_0\;|\sin{l}|) \\
    & & \;\; -\,
    (v_\mathrm{c}(R_0)+V_\odot)\sin{l} - U_\odot \cos{l}. \\
  \end{array}
\end{equation}
with uncertainty of $7\kms$ (N.B. the function ${\rm sgn}(x)$ is simply $1$ if $x>0$ and $-1$ if $x<0$). 
This assumed uncertainty is to make allowances for the
effects that non-axisymmetric structure in the Galaxy and non-circular
motion of the ISM will have on these data.

\subsubsection{Vertical force} \label{sec:kz}
We use the \cite{KuGi91} value for the vertical force
at $1.1\kpc$ above the plane at the Solar radius, $K_{z,1.1,\odot}$ as a constraint.
\begin{equation}
  K_{z,1.1,\odot} = 2\pi G \times (71\pm6)\msun\pc^{-2}.
\end{equation}
It is worth noting \citep[as in the review by][]{Re14} that \citeauthor{KuGi91} do apply a loose rotation curve prior to their analysis, so there is some danger of `double counting' when using this constraint along with data constraining the rotation curve. However, the \cite{KuGi91} prior is very broad, and only affects the values of the parameters that compare the contributions of the disc and halo. Since the value of $K_{z,1.1,\odot}$ derived by \citeauthor{KuGi91} appears to be almost independent of these parameters, we feel confident using this constraint.

\subsubsection{Mass within large radii}
It remains tremendously difficult to constrain the mass of the Milky Way out to large radii. In Section~\ref{sec:others} we discuss recent efforts. We again adopt a constraint inspired by \cite{WiEv99} on the
mass within the Galaxy's inner $50\kpc$, which we refer to as $M_{50}$. 
\begin{equation}\label{eq:m50}
  P(M_{50})= \left\{ 
    \begin{array}{ll}
      C & \mathrm{for }\,M_{50} \leq M_{\mathrm{WE}} \\
      C\exp\left(-\left[\frac{M_{50}-M_{\mathrm{WE}}}{\delta_{M_{50}}}
        \right]^2\right) &\mathrm{for }\,  M_{50} > M_{\mathrm{WE}} \\
    \end{array}
  \right.
\end{equation}
where $M_{\mathrm{WE}}=5.4\times10^{11}\msun$,
$\delta_{M_{50}}=2\times10^{10}\msun$, and $C$ is a normalisation
constant. This effectively sets an upper limit on $M_{50}$. In section~\ref{sec:others} we discuss other studies of the outer reaches of the Milky Way, and how they compare to our results. 

\section{Fitting the models} \label{sec:fit} 

In Paper I we outlined a scheme to fit mass models of the Galaxy to
the sort of constraints described above. We use the same techniques
again here. They were also used by \cite{Piea14} as constraints on
dynamical models fit to RAVE data \citep{RAVEDR4}, though in that study
only best fitting models were found in each case, not the full pdf on
all model parameters.

We refer to the parameters collectively as $\bolth$, and the data as
$d$. Bayes theorem tells us that this pdf is then
\begin{equation}
  p(\bolth|d) = \frac{\mathcal{L}(d|\bolth)\,p(\bolth)}{p(d)}
\end{equation}
where the total likelihood $\mathcal{L}(d|\bolth)$ is the product of
the likelihoods associated with each kinematic data-set or constraint
described in Section~\ref{sec:data}, and $p(\bolth)$ is the
probability of a parameter set given the prior probability
distributions described in Section~\ref{sec:comp}. The `evidence',
$p(d)$, is simply a normalisation constant in this study, and we
therefore ignore it.

We have 15 free parameters: The scale lengths and density
normalisations of the thin and thick discs
($R_{\d,\mathrm{thin}},\Sigma_{0,\mathrm{thin}},
R_{\mathrm{d},\mathrm{thick}},\Sigma_{0,\mathrm{thick}}$); the density
normalisation -- and thus mass -- of the bulge
($\rho_{0,\mathrm{b}}$); the scale radius and density normalisation of
the CDM halo ($r_\mathrm{h},\rho_{0,\mathrm{h}}$); the solar radius
($R_0$); the three components of $\vsol$; the three components of $\vsfr$,
and the typical random component of the HMSFR velocity $\Delta_v$. We
have direct priors on many of these parameters, or on quantities that
can be directly derived from them, as described above.

Clearly this is too large a parameter space to explore by brute
force. We therefore use a Markov Chain Monte Carlo (MCMC) method, the
Metropolis algorithm \citep{Metropolis}, to explore the pdf. This
explores the full pdf, and the output chain is a fair sample of the pdf. As a test, we have also used the affine invariant MCMC sampler proposed by \cite{GoWe10}, and packaged as \textsc{emcee} by \cite{emcee}, to explore the pdf of some of our models, and found essentially identical results.

\section{Main model}\label{sec:Main}
\begin{table*}
  \begin{tabular}{ccccccccc} 
    \hline
     & $\Sigma_{0,\mathrm{thin}}$ & $R_{\d,\mathrm{thin}}$ & 
    $\Sigma_{0,\mathrm{thick}}$ & $R_{\d,\mathrm{thick}} $ & 
    $\rho_{0,\mathrm{b}}$ & $\rho_{0,\mathrm{h}}$ & $r_\mathrm{h} $ & 
    $R_0 $  \\ \hline
%
 \rule[-1.2ex]{0pt}{0pt} Mean $\pm$ Std Dev &  $ 886.7 \pm 116.2 $ & $ 2.53 \pm 0.14 $ & $ 156.7 \pm 58.9 $ & $ 3.38 \pm 0.54 $ & $ 97.3 \pm 9.7 $ & $ 0.0106 \pm 0.0053 $ & $ 19.0 \pm 4.9 $ & $ 8.20 \pm 0.09 $ \\  
Median \& range & $ 887.0 _{- 115.0 }^{+ 116.6 }$ & $ 2.53 _{- 0.14 }^{+ 0.15 }$ & $ 148.7 _{- 53.2 }^{+ 73.1 }$ & $ 3.29 _{- 0.45 }^{+ 0.63 }$ & $ 97.3 _{- 9.8 }^{+ 9.7 }$ & $ 0.0093 _{- 0.0034 }^{+ 0.0059 }$ & $ 18.6 _{- 4.4 }^{+ 5.3 }$ & $ 8.20 _{- 0.09 }^{+ 0.09 }$ \\ \hline 
  & $U_\odot$ & $V_\odot$ & $W_\odot$ & $v_{R,{\rm SFR}}$ &
   $v_{\phi,{\rm SFR}}$ & $v_{z,{\rm SFR}}$ & $\Delta_v$ & \\\hline
\rule[-1.2ex]{0pt}{0pt} Mean $\pm$ Std Dev &  $ 8.6 \pm 0.9 $ & $ 13.9 \pm 1.0 $ & $ 7.1 \pm 1.0 $ & $ -2.7 \pm 1.4 $ & $ -1.1 \pm 1.3 $ & $ -1.9 \pm 1.4 $ & $ 6.8 \pm 0.6 $ & \\
Median \& range & $ 8.6 _{- 0.9 }^{+ 0.9 }$ & $ 13.9 _{- 1.0 }^{+ 1.0 }$ & $ 7.1 _{- 1.0 }^{+ 1.0 }$ & $ -2.7 _{- 1.4 }^{+ 1.4 }$ & $ -1.1 _{- 1.3 }^{+ 1.3 }$ & $ -1.9 _{- 1.4 }^{+ 1.4 }$ & $ 6.8 _{- 0.6 }^{+ 0.6 }$ &  \\ \hline
& $v_0$ & $M_{\mathrm{b}}$ & $M_\ast$  & $M_v $ & $c_{v'}$ &
   $K_{z,1.1,\odot}$ & $\rho_{\mathrm{h},\odot}$ & \\ \hline
\rule[-1.2ex]{0pt}{0pt} Mean $\pm$ Std Dev &  $ 232.8 \pm 3.0 $ & $ 9.13 \pm 0.91 $ & $ 54.3 \pm 5.7 $ & $ 1300 \pm 300 $ & $ 16.4 \pm 3.1 $ & $ 74.8 \pm 4.9 $ & $ 0.0101 \pm 0.0010 $ &  \\  
Median \& range & $ 232.8 _{- 3.0 }^{+ 3.0 }$ & $ 9.13 _{- 0.92 }^{+ 0.91 }$ & $ 54.3 _{- 5.8 }^{+ 5.6 }$ & $ 1300 _{- 300 }^{+ 300 }$ & $ 16.0 _{- 2.7 }^{+ 3.4 }$ & $ 74.8 _{- 4.9 }^{+ 4.9 }$ & $ 0.0101 _{- 0.0010 }^{+ 0.0010 }$ & \\
 \end{tabular}
  \caption{
    Expectation values and uncertainties (upper rows), and median and $\pm1\sigma$ equivalent range (lower rows) for the free parameters of the
    mass model as described by eqs.~\ref{eq:bulge}, \ref{eq:disc} and
    \ref{eq:halo} (top), the peculiar velocity of the Sun and free 
    parameters of the maser velocity pdf as described by
    eqs.~\ref{eq:vbar} and \ref{eq:veldist} (middle), and derived
    properties of the mass model (bottom). $M_{\mathrm{b}}$ is the
    bulge mass, $M_\ast$ is the total stellar mass, and
    $\rho_{\mathrm{h},\odot}$ is the halo density at the Sun's
    position. 
    Distances are quoted in units of $\mathrm{kpc}$, velocity in $\mathrm{km\,s}^{-1}$, 
    masses in $10^9\msun$, surface densities in $\msun\pc^{-2}$,
    densities in $\msun\pc^{-3}$, and $K_{z,1.1,\odot}$ in units of 
    $(2\pi G)\times\msun\pc^{-2}$. The local dark-matter density can
    also 
    be written as $(0.38\pm 0.04)\Gevdens$. 
    \label{tab:mainresults}
  }
\end{table*}

Our `main' model has an NFW halo (i.e. $\gamma=1$ in
eq~\ref{eq:halo}). In Table~\ref{tab:mainresults} we give the expectation values and
standard deviations, and median and $\pm1\sigma$ equivalent range, for the parameters of our model, and various derived
quantities of interest. We define the  $\pm1\sigma$ equivalent range as being between the $15.87$ and $84.13$ per cent points in the cumulative distribution (the percentage equivalents of the 1$\sigma$ range in a Gaussian distribution). We include this because a few of our parameters have significantly non-Gaussian distributions, but in most cases we will quote the standard deviations of values.

\begin{table}
  \begin{tabular}{cc|cc}
    Parameter &  & Property  & \\\hline
    $\Sigma_{0,\mathrm{thin}}$ & $896\msun\pc^{-2}$     & $v_0$ & $233.1\kms$ \\
    $R_{\d,\mathrm{thin}}$         &  $2.50\kpc$      & $M_b$ & $9.23\times10^9\msun$ \\
    $\Sigma_{0,\mathrm{thick}}$ & $183\msun\pc^{-2}$    & $M_\ast$ & $5.43\times10^{10}\msun$ \\
    $R_{\d,\mathrm{thick}} $       & $3.02\kpc$      & $M_v$ & $1.37\times10^{12}\msun$ \\
    $\rho_{0,\mathrm{b}}$           & $98.4\msun\pc^{-3}$    & $c_{v'}$ & $15.4$ \\
    $\rho_{0,\mathrm{h}}$          & $0.00854\msun\pc^{-3}$ & $K_{z,1.1,\odot}$ & $73.9\times(2\pi G)\msun\pc^{-2}$ \\
    $r_\mathrm{h} $                   & $19.6\kpc$        & $\rho_{\mathrm{h},\odot}$ & $0.0101\msun\pc^{-3}$ \\
    $R_0 $                                 & $8.21\kpc$ & & \\
  \end{tabular} \caption{
    Parameters and properties of our best fitting model.
  }\label{tab:bestmodel}
\end{table}

In Table~\ref{tab:bestmodel} we give the
parameters of our best fitting model. Figures~\ref{fig:Rdisc} to
\ref{fig:M} show the marginalized pdfs for various properties of the
models.

\begin{table*}
  \begin{tabular}{c|ccccccccc}
    &  $\Sigma_{0,\mathrm{thin}}$ & $R_{\d,\mathrm{thin}}$ & 
    $\Sigma_{0,\mathrm{thick}}$ & $R_{\d,\mathrm{thick}} $ & 
    $\rho_{0,\mathrm{b}}$ & $\rho_{0,\mathrm{h}}$ & $r_\mathrm{h} $ & 
    $R_0 $  \\ \hline
    $\Sigma_{0,\mathrm{thin}}$ & 1 &  &  &  &  &  &  &  \\ 
    $R_{\d,\mathrm{thin}}$ & -0.49 & 1 &  &  &  &  &  &  \\ 
$\Sigma_{0,\mathrm{thick}}$ & -0.13 & 0.27 & 1 &  &  &  &  &  \\ 
$R_{\d,\mathrm{thick}} $ & 0.23 & -0.05 & -0.84 & 1 &  &  &  &  \\ 
$\rho_{0,\mathrm{b}}$  & -0.43 & 0.30 & -0.12 & 0.09 & 1 &  &  &  \\ 
$\rho_{0,\mathrm{h}}$ & -0.61 & 0.02 & -0.42 & 0.25 & 0.04 & 1 &  &  \\ 
$r_\mathrm{h} $ & 0.58 & -0.04 & 0.43 & -0.25 & -0.06 & -0.90 & 1 &  \\ 
$R_0 $ & -0.14 & 0.38 & -0.01 & 0.10 & -0.02 & 0.14 & -0.11 & 1 \\ 
%
%
  \end{tabular} \caption{
    Correlations between the parameters describing the mass model of the Milky Way (and the Sun's position with in it). The values are those given by eq.~\ref{eq:corr}. The full two dimensional pdfs that these values describe are also plotted in Fig.~\ref{fig:Corner}.
  }\label{tab:coo_par}
\end{table*}

\begin{figure*}
  \centerline{\resizebox{\hsize}{!}{\includegraphics{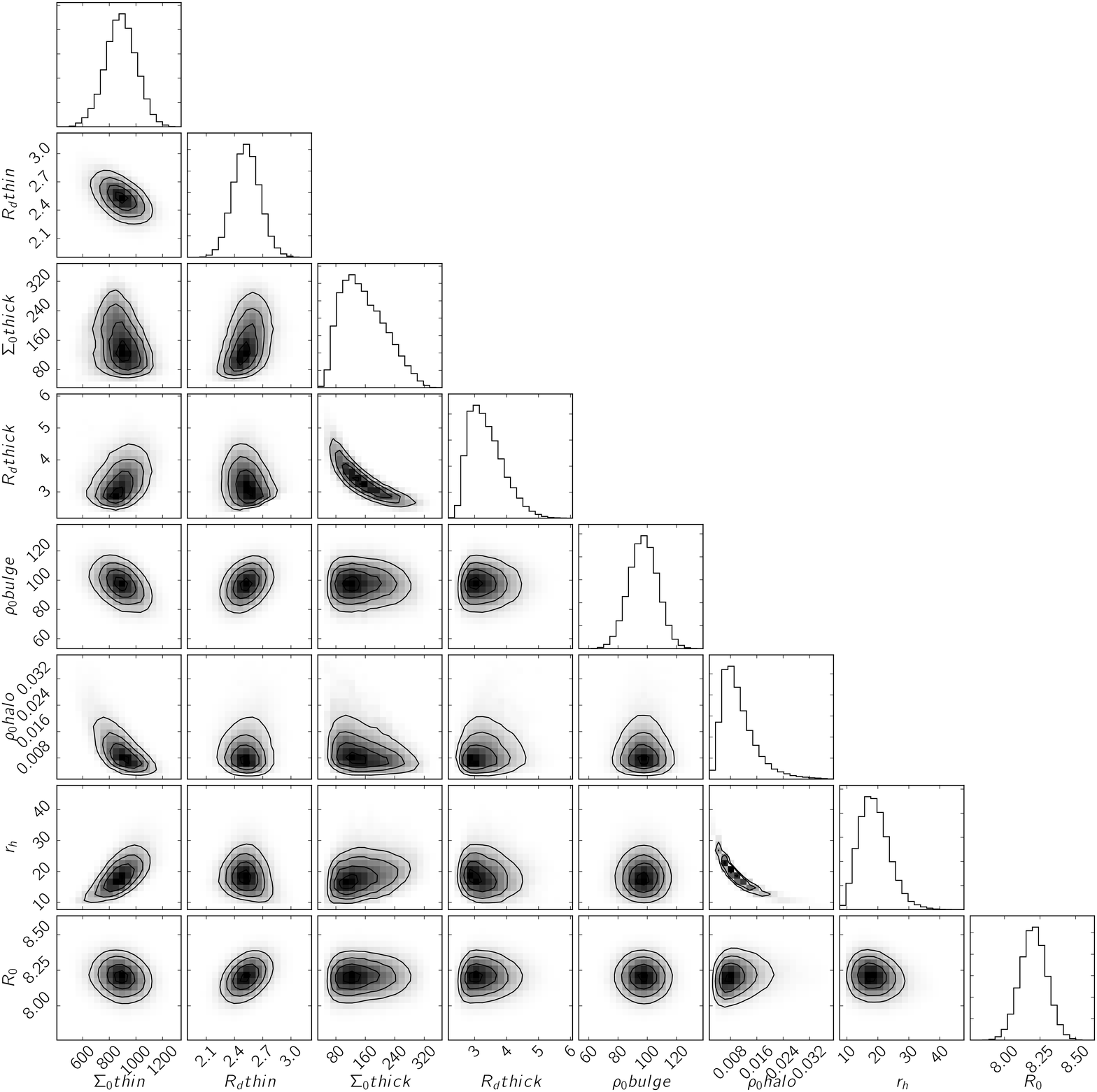}}}
  \caption{2-dimensional pdf of the main parameters. The correlation coefficients are given in Table~\ref{tab:coo_par}. 
\label{fig:Corner}
}
\end{figure*}

\begin{figure}
  \centerline{\resizebox{\hsize}{!}{\includegraphics[angle=270]{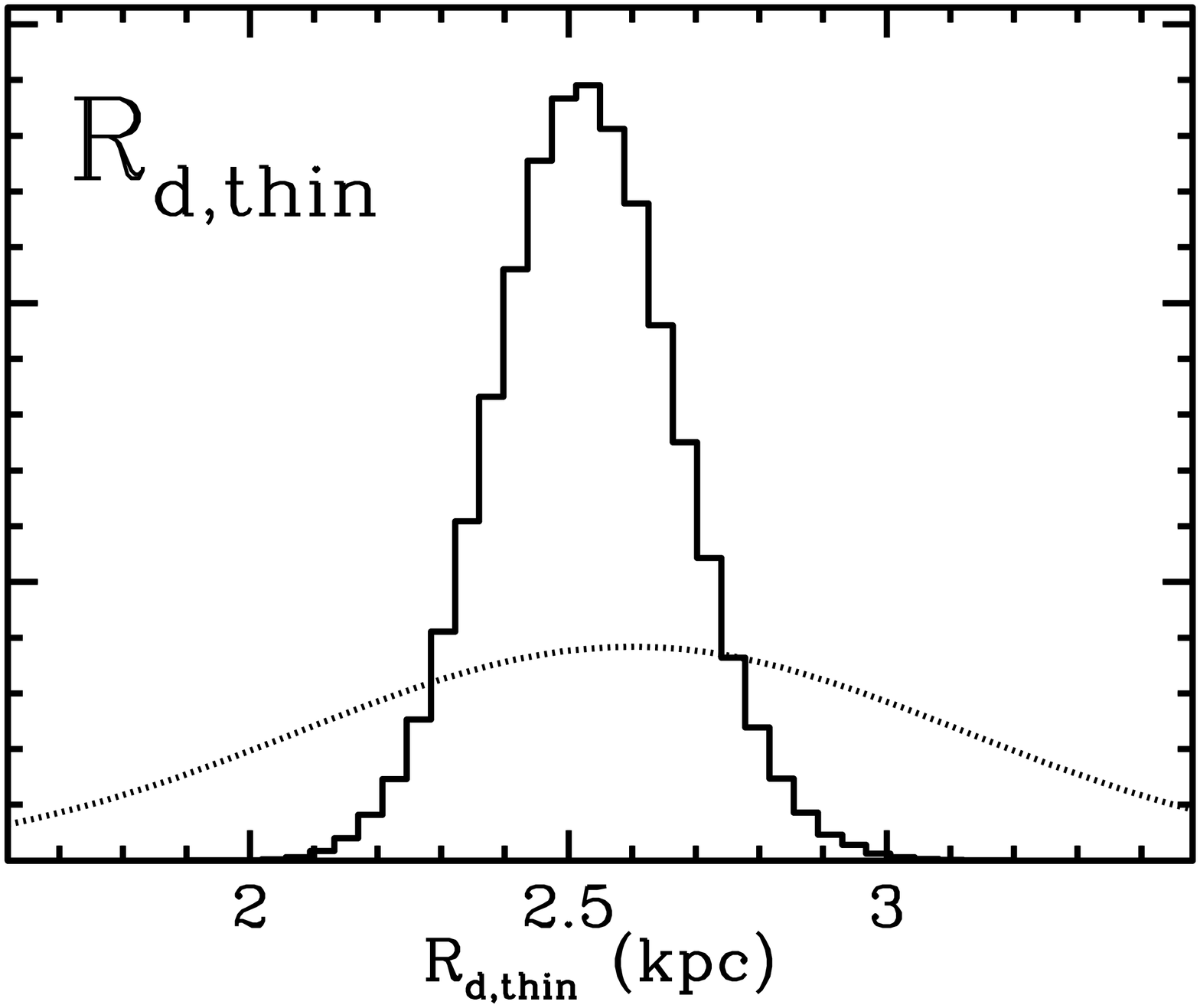}}}
  \caption{Histogram of the pdf of the thin-disc 
    scale lengths in our model (solid histogram) normalised over all other
    parameters, compared in each case to the prior 
    pdf described in Section~\ref{sec:stellardisc} (dotted).
\label{fig:Rdisc}
}
\end{figure}

\begin{figure}
  \centerline{\resizebox{0.8\hsize}{!}{\includegraphics[angle=270]{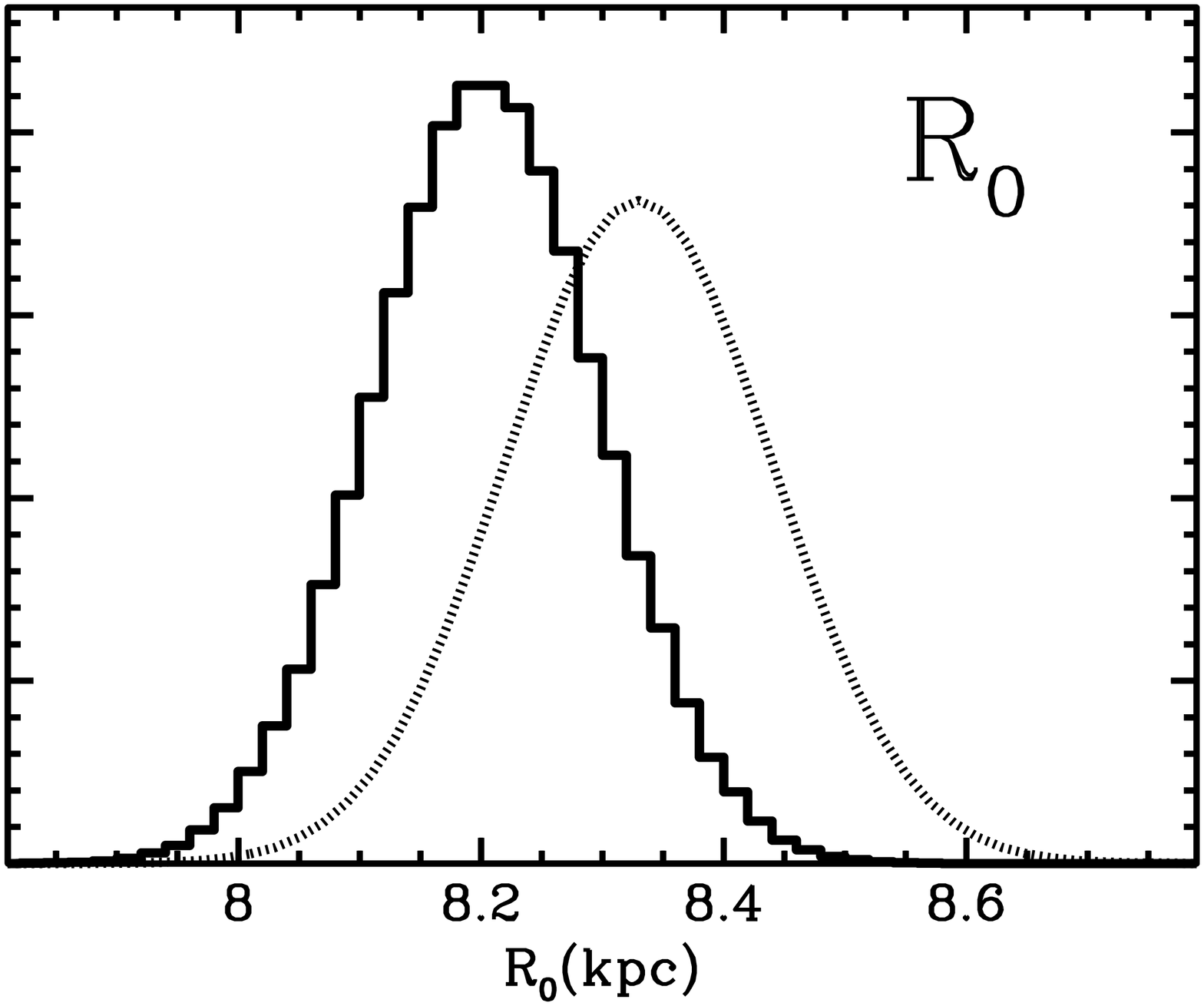}}}\vspace{-2mm}
  \centerline{\resizebox{0.8\hsize}{!}{\includegraphics[angle=270]{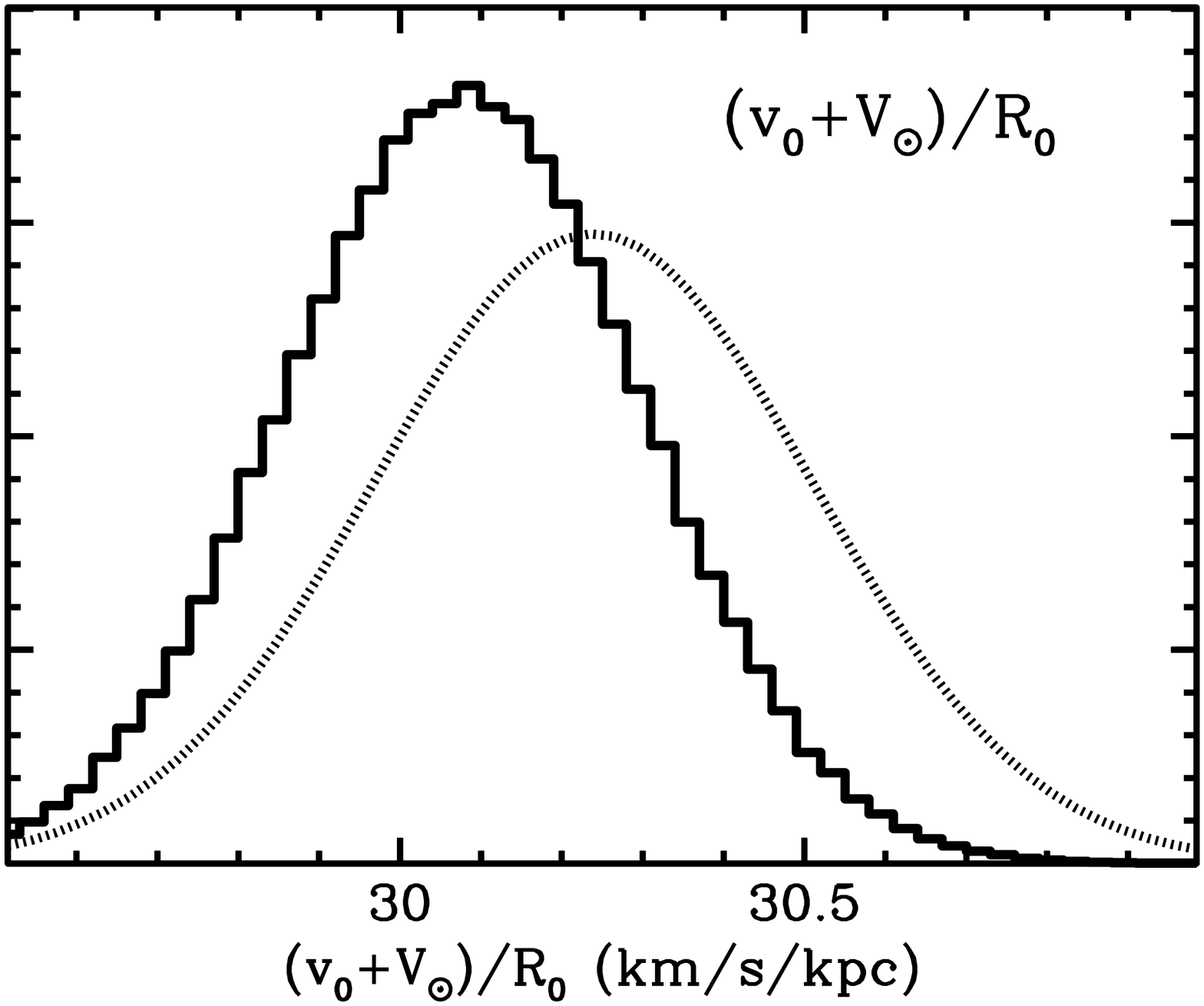}}}
  \caption{Histograms of the pdf of $R_0$ (upper) and
    $(v_0+V_\odot)/R_0$ (lower) with the model output shown as a
    solid histogram, and the prior on each quantity shown as a dotted
    curve. In the lower panel the prior shown is that associated with
    the proper motion of Sgr A*. 
\label{fig:R0v0}
}
\end{figure}

\begin{figure}
  \centerline{\resizebox{0.5\hsize}{!}{\includegraphics[angle=270]{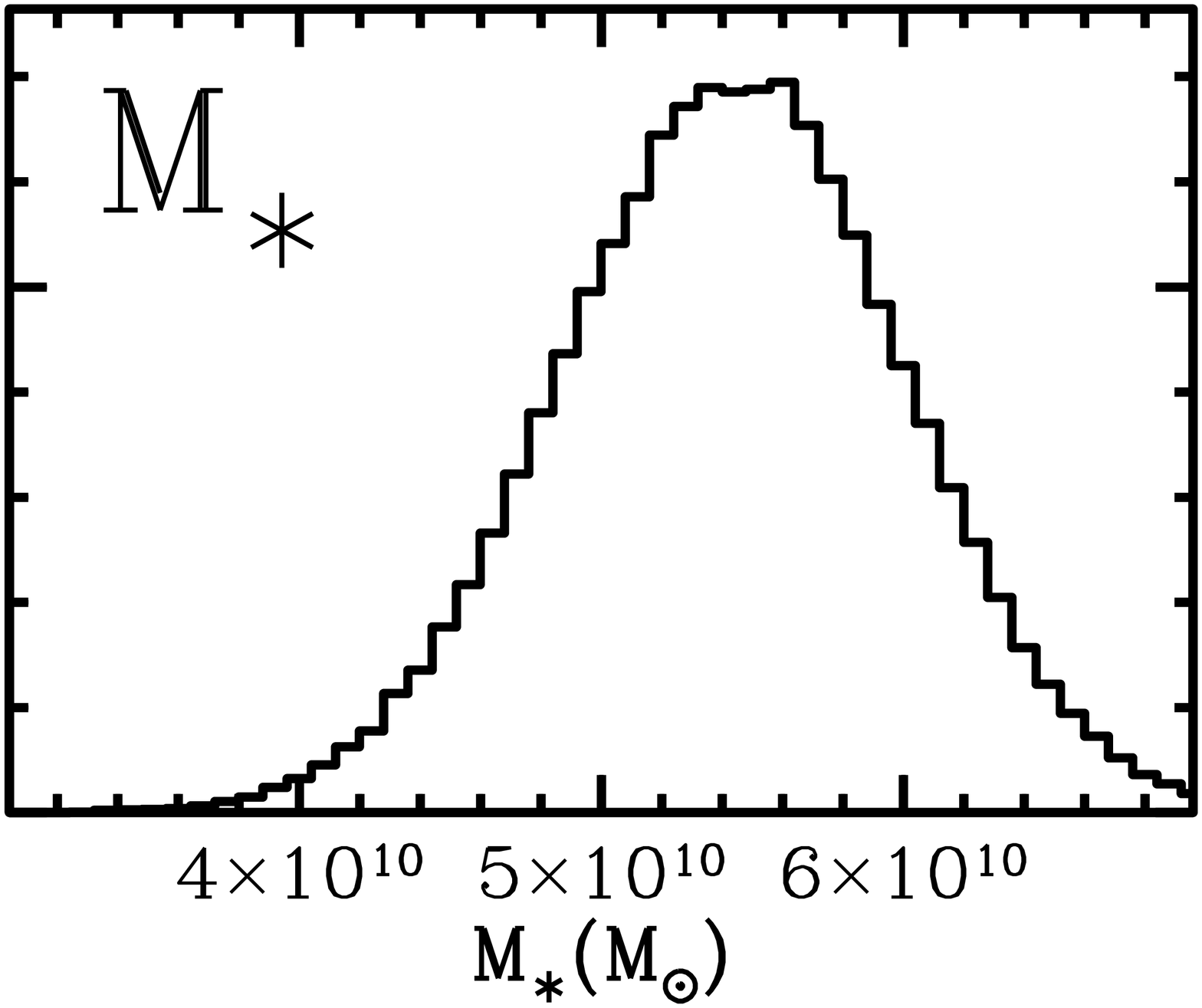}}
    \resizebox{0.5\hsize}{!}{\includegraphics[angle=270]{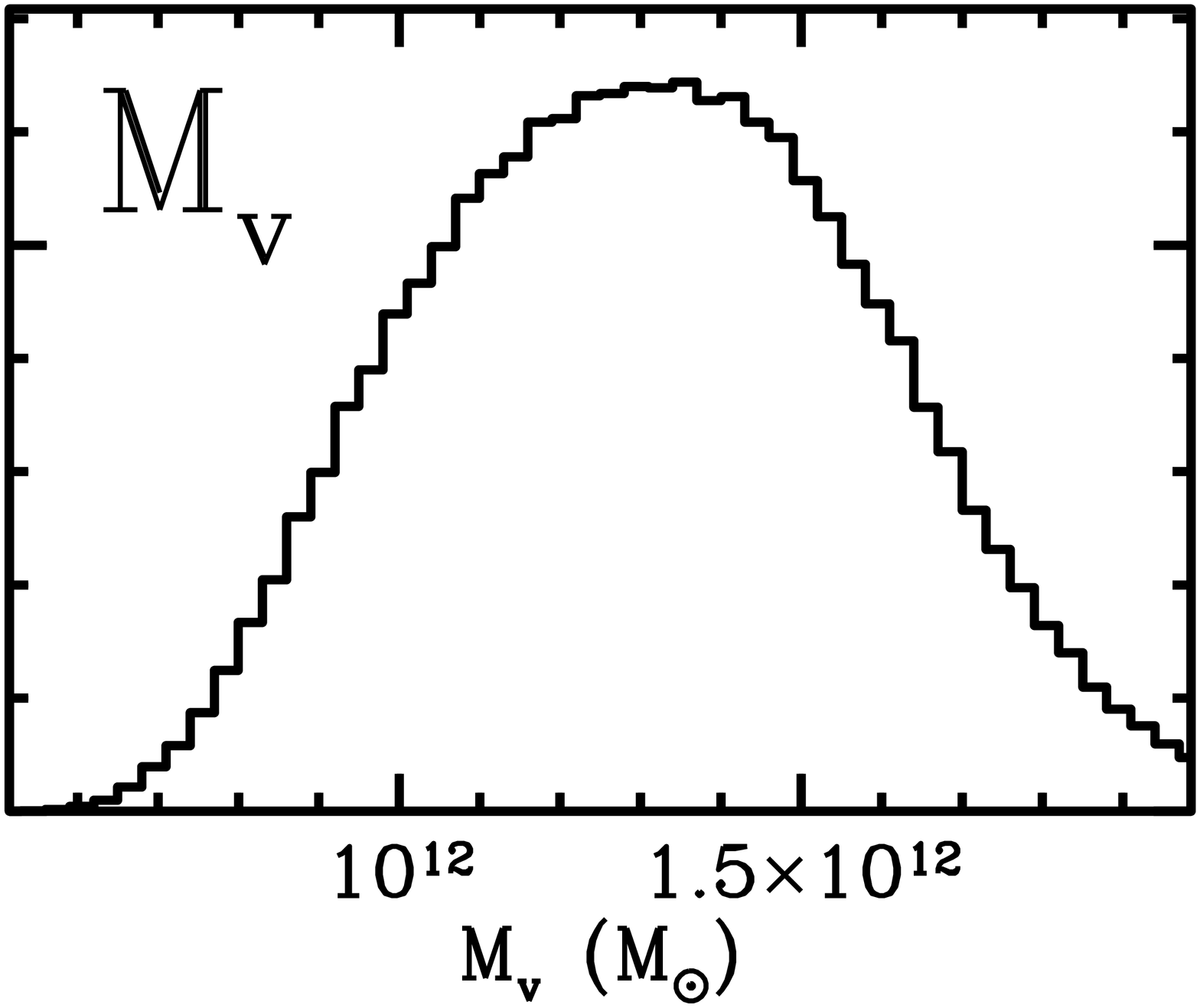}}}
  \vspace{-1mm}
  \centerline{\resizebox{0.5\hsize}{!}{\includegraphics[angle=270]{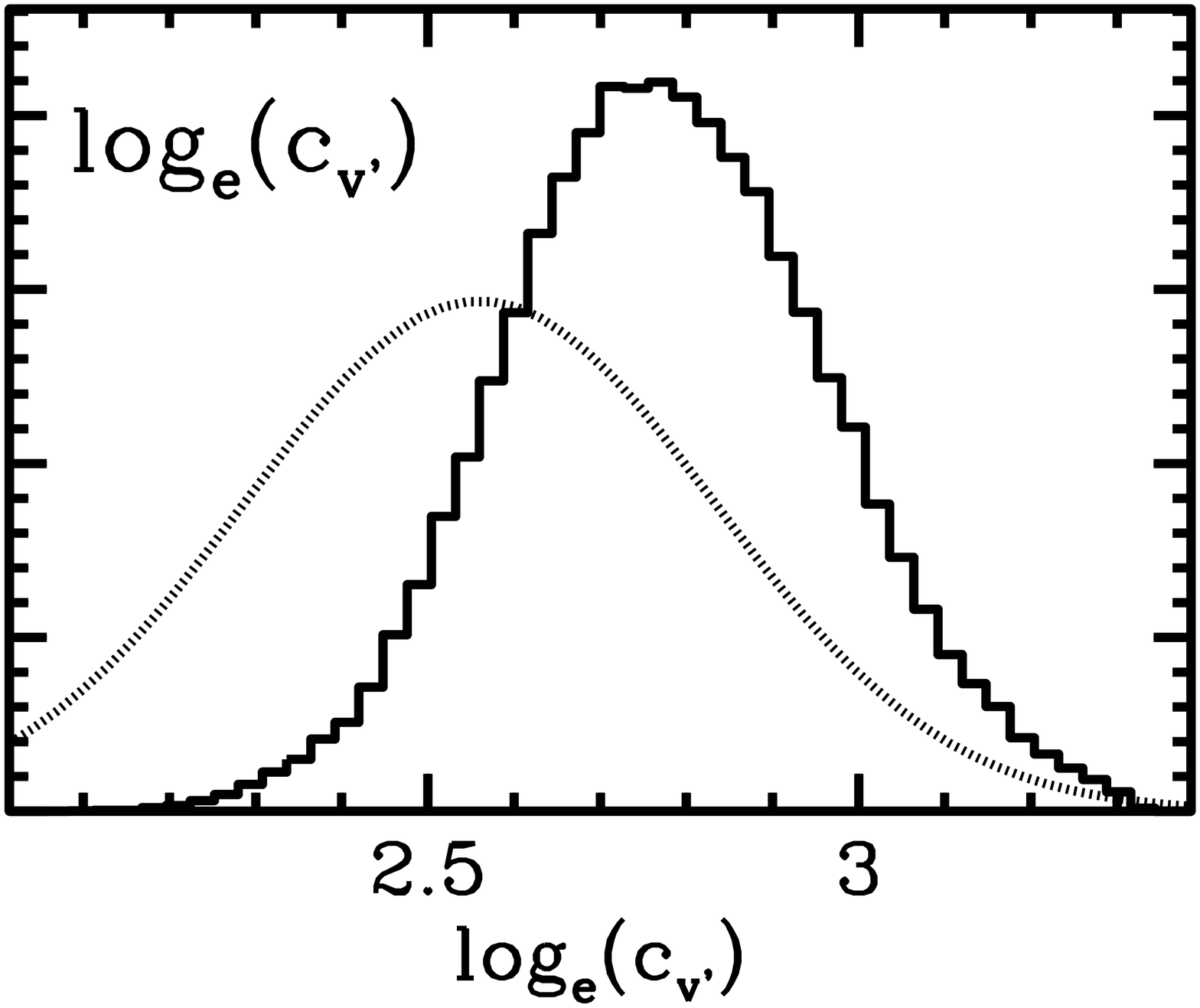}}
    \resizebox{0.5\hsize}{!}{\includegraphics[angle=270]{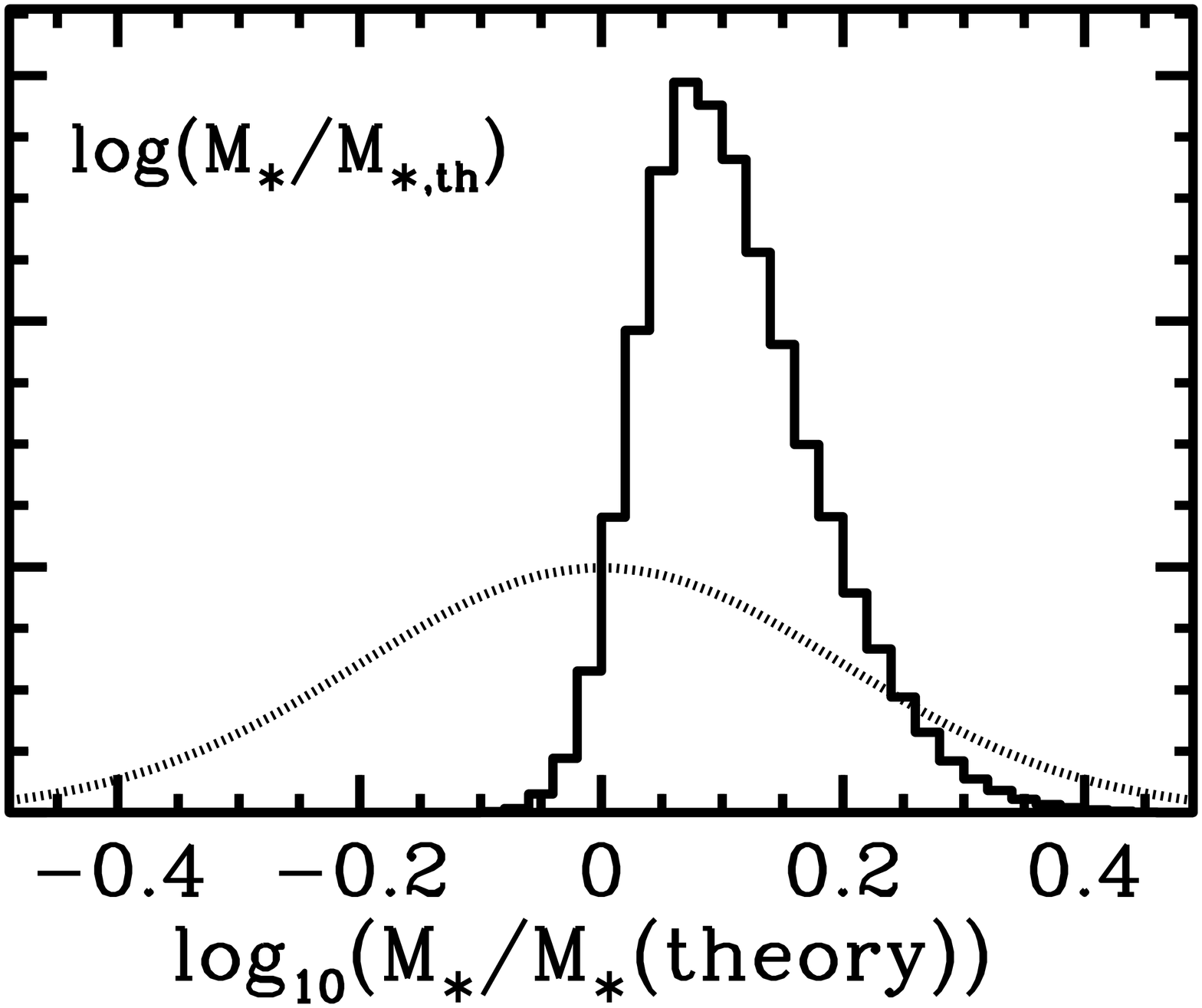}}}
  \caption{Histograms of the pdf total stellar mass ($M_\ast$,
    \emph{upper-left}), virial mass ($M_v$, \emph{upper-right}), natural logarithm
    of the halo concentration ($\log c_{v'}$, \emph{lower-left}), and the
    logarithm of the ratio of
    the stellar mass to that predicted by eq.~\ref{eq:mast} given the
    virial mass (\emph{lower-right}). In the lower panels the priors
    on the plotted properties are shown as dotted curves.
\label{fig:M}
}
\end{figure}

These 1D marginalised pdfs do not tell the full story, of course, because many of the parameters are correlated with one another. In Fig.~\ref{fig:Corner} we show a `corner plot', which is a density/scatter plot of 2D projections of the pdf that shows the relationships between the parameters which define the mass model (and $R_0$). For a more quantitative representation we also give the correlation matrix for these parameters. For parameters $\vtheta$, this is made up of the values 
\begin{equation} \label{eq:corr}
  \mathrm{corr}(\theta_i,\theta_j) = \frac{\mathrm{cov}(\theta_i,\theta_j)}
  {\sigma_i\sigma_j},
\end{equation}
where $\mathrm{cov}(\vv_i,\vv_j)$ is the covariance of the two
parameters, and $\sigma_i$ and $\sigma_j$ their standard
deviation. It takes values between $-1$ and $1$, with $1$ being
perfectly correlated, $-1$ being perfectly anti-correlated and $0$
being uncorrelated. 

The strongest correlations or anti-correlations are typically, as in Paper I, between parameters that combine to define the properties (such as the total mass) of a component. $\rho_{0,\mathrm{h}}$ and $r_{\mathrm{h}}$ are very strongly anti-correlated, as are $\Sigma_{0,\mathrm{thick}}$ and $R_{\d,\mathrm{thick}}$. This explains, for example, why the spread in $\rho_{0,\mathrm{h}}$ is much larger than that in $M_v$. There are fairly strong correlations between $\Sigma_{0,\mathrm{thin}}$ and most other parameters, and between $R_0$ and $R_{\d,\mathrm{thin}}$. This is very similar to the correlations found in Paper I.

Fig.~\ref{fig:Rdisc} shows the pdf for the scale length of the thin disc, along with the prior from \cite{Juea08}. The peaks lie at nearly the same value of $R_{\d,\mathrm{thin}}$, but the range of plausible values we find is smaller than in the prior. This is noticeably different from the value of $(3.00\pm0.22)\kpc$ found in Paper I. This is partially due to the new data we consider, and partly due to the omission of a gas disc in Paper I. If we omit the gas discs from our models, we find thin-disc scale lengths of ${\sim}2.8\kpc$. The gas discs reduce the value of $R_{\d,\mathrm{thin}}$ because the \HI\ disc provides a component of the disc mass that has a longer scale length than the stellar components. To compensate in the fit of the kinematic data, the scale length of the stellar component decreases. This emphasises the importance of modelling assumptions in finding the parameters, and therefore of including a component to represent the gas disc. In Section~\ref{sec:hiloGas} we explore the consequences of changing the mass of the gas component in more detail.

The pdfs of $R_0$ and $(v_0+V_\odot)/R_0$, shown in Fig.~\ref{fig:R0v0}, are both displaced to slightly lower values than the priors (which, in the latter case, is the proper motion of Sgr A*). In the case of $R_0$, this is ${\sim}1\sigma$ lower than the value from the prior. The value of $R_0$ is clearly pulled down by the need to fit the maser data (a effect that was not noticeable when using data from the $24$ masers observed before Paper I). In experiments where we applied a weaker prior to $R_0$ ($8.33\pm0.35 \kpc$, as in Paper I), the value of $R_0$ found is even lower, at ${\sim}8.0\kpc$.

This lower value of $R_0$ is the primary cause of our derived value of $v_0$ being lower than in Paper I (where it was ${\sim}239\kms$). Our value of $232.8\kms$ lies neatly between the `traditional' value of $220\kms$, and the values found by more recent studies which tend to be closer to ${\sim}240\kms$ \citep[e.g.][]{Sc12,Reea14}. It is rather close to the value found by \cite{Shea14} from RAVE data.

In Fig. \ref{fig:M} we show pdfs of the stellar mass and virial mass of our models (upper panels). The lower panels show how the halo concentration compares to our prior (typically slightly higher than expected, but well within the uncertainties), and how the stellar mass compares to our prior, given the virial mass (again, slightly higher than expected, but well within the range expected). These results are very similar to those found in Paper I.

\begin{figure}
  \centerline{\resizebox{0.7\hsize}{!}{\includegraphics[angle=270]{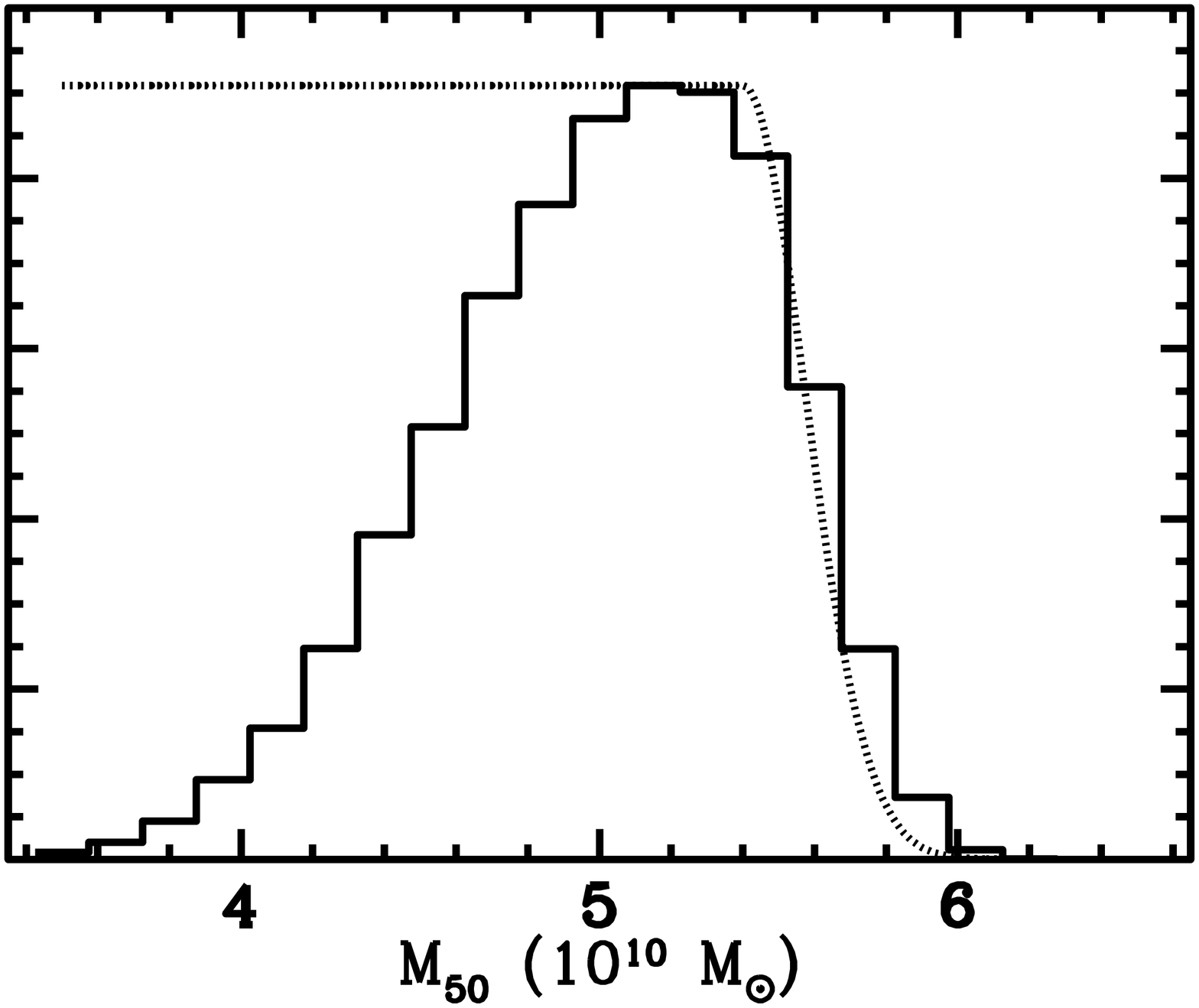}}}
  \caption{Pdf of the mass contained within $50\kpc$ ($M_{50}$, solid histogram). The dotted line illustrates the prior on $M_{50}$. Since this prior is a constant for $M_{50}<5.4\times10^{11}$, we have artificially chosen to rescale this constant to the same value as the peak of the binned pdf in the interests of making comparison easier.
\label{fig:M50}
}
\end{figure}

In Figure~\ref{fig:M50} we show the pdf of values for $M_{50}$, with a representation of the prior for comparison. The prior clearly provides an important upper bound. This is to be expected, because no other data that we use as input is able to provide any useful constraint on the potential in the outer parts of the Galaxy. We observed the same behaviour in Paper I. It is clear that there are a range of values of $M_{50}$ below $M_{\mathrm{WE}}$ that provide acceptable models. If we remove the prior on $M_{50}$, then the peak of our pdf in $M_{50}$ is still found at $\sim M_{\mathrm{WE}}$, but the distribution is more symmetric about the peak.

\subsection{Maser velocities}
The typical peculiar velocity associated with the maser sources is
small, the largest being a velocity of $(2.7\pm 1.3)\kms$ radially
inwards. We find that the claimed lag in the rotational velocity is
negligible: $(1.1\pm1.4)\kms$.  This is in contrast to the 
lag that is claimed by \cite{Reea14} using the same data,
which was $(5.0\pm2.2)\kms$ when they use the \cite{SBD10} value of
$\vsol$ as a prior. There are three main differences between this study
and that one that, together, are likely to have caused this
difference. Firstly, that our rotation curves comes from our mass model,
rather than any of the assumed forms used by
\citeauthor{Reea14}. Secondly, that we marginalize over all possible
distances to the objects, rather than approximating that the quoted
parallax gives the true distance. We note that in some cases (e.g. G023.65-00.12 and G029.95-00.01) this means that sources labelled as outliers by \cite{Reea14} prove to be well fit by our models within the parallax uncertainties but not at the quoted parallax. Thirdly, we use an outlier model to
take account of objects which have significantly non-circular motions,
rather than removing them from the analysis entirely. These results
are not significantly altered if we increase the outlier fraction
$f_{\rm out}$ to $0.1$.

The intrinsic spread of the HMSFR velocities around a circular orbit
(ignoring the outliers) is $(6.8\pm 0.6)\kms$, which is comparable to the
values found by \cite{PJMJJB10:Masers} and the value assumed in Paper I. It
is a little larger than the value of $5\kms$ assumed by \cite{Reea14}.

\subsection{Solar position and velocity} \label{sec:solarv}
\begin{figure}
  \centerline{\resizebox{\hsize}{!}{\includegraphics{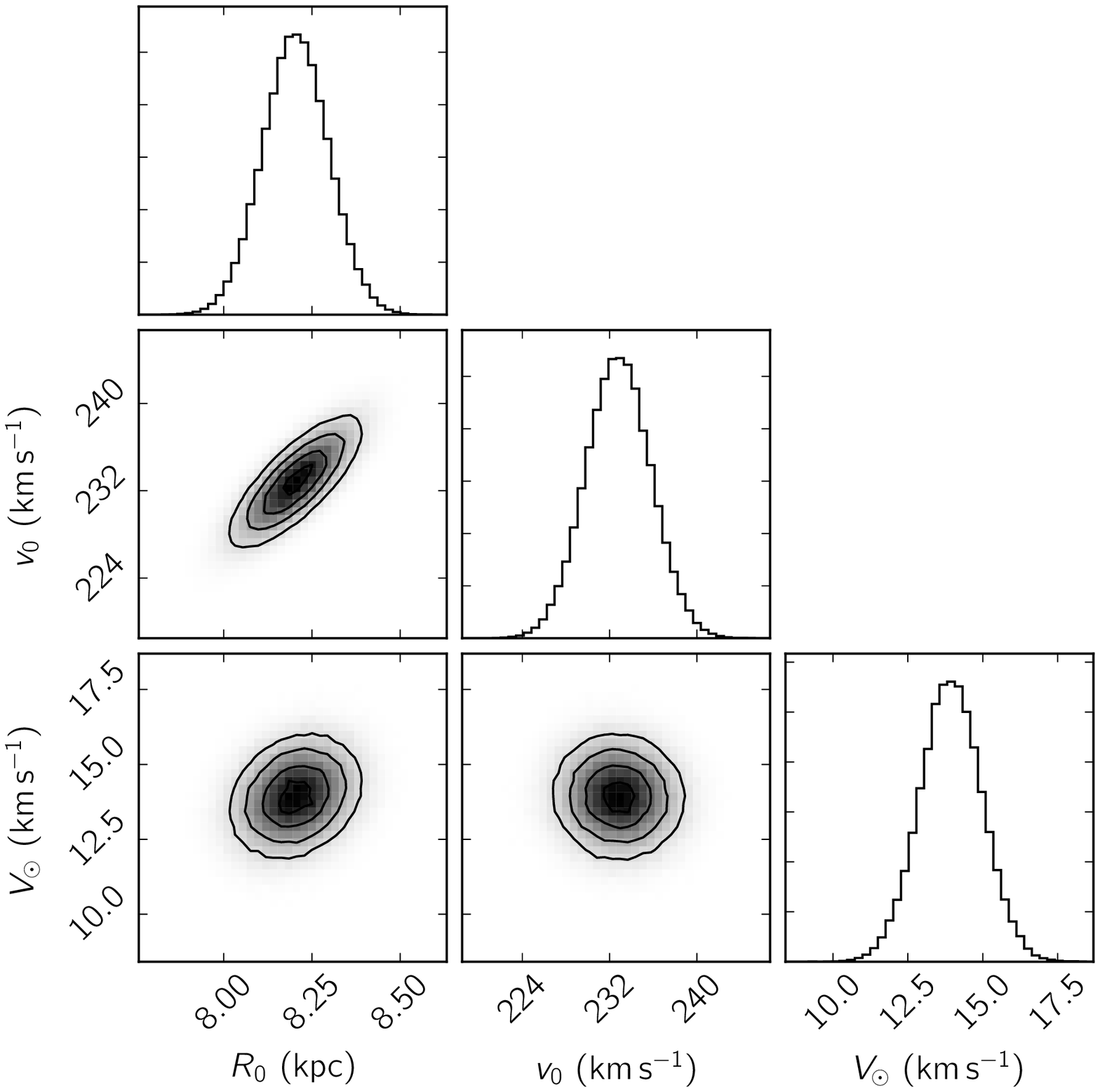}}}
  \caption{Two dimensional histograms showing the distribution of values found in the MCMC simulation for $R_0$, $v_0$ and $V_\odot$. The dominant feature is the correlation between $R_0$ and $v_0$.
\label{fig:RvV}
}
\end{figure}

The value of $R_0$ we find is $(8.20 \pm 0.09)\kpc$, which is ${\sim}
1\sigma$ below the value taken as a prior (eq.~\ref{eq:R0}). It is
also a little lower than the value found in Paper I of $(8.29\pm0.16)\kpc$, by
\cite{Sc12} of $(8.27\pm 0.29)\kpc$ or by \cite{Reea14} of
$(8.34\pm0.16) \kpc$. It is still consistent with these values, and
with the distance to the Galactic centre found from the parallax to
Sgr B2 \citep[$7.9^{+0.8}_{-0.7}\kpc$, ][]{Reea09:SgrB2}. It is very close to the distance (and uncertainty) found from a compilation of literature values by \cite{BHGe16}. Since the prior placed on $v_0/R_0$ by
the proper motion of Sgr A* is still closely followed, we therefore
find a somewhat lower value of $v_0$ than Paper I, of
$(232.6\pm 3.0)\kms$. This reemphasises the point made by
\cite{PJMJJB10:Masers} that estimates of $v_0$ are typically highly dependent on $R_0$, so it is dangerous to treat $R_0$ as known and fixed. In Figure~\ref{fig:RvV} we plot the distribution of values of $R_0$, $v_0$ and $V_\odot$ from our MCMC chain. The clearest feature is the strong correlation between $R_0$ and $v_0$.

The differences between these values of $R_0$ and $v_0$ and those found
in Paper I are at around the $1\sigma$ level (less when both
uncertainties are taken into account). These can be attributed to the
changed modelling assumptions (including the addition of a gas disc
and allowing $\vsol$ to vary) as well as the additional data.

The peculiar velocity $\vsol$ that we find differs
somewhat from our prior taken from \cite{SBD10}, though only at the
$1$ to $2\sigma$ level. Our results suggest that the value of $U_\odot$ may
be a little smaller than suggested by \citeauthor{SBD10}, and
$V_\odot$ may be a little larger.

The values of the components of $\vsol$ are correlated with
the components of the typical peculiar motion associated with the maser sources
$\vsfr$. Table~\ref{tab:sun_maser} shows the correlation matrix of
the components of $\vsol$ with the components of $\vsfr$.

\begin{table}
  \begin{tabular}{c|ccc}
    & $U_\odot$ & $V_\odot$ & $W_\odot$ \\\hline
    $v_{{\rm SFR},R}$ & -0.57 & 0.00 & -0.03 \\
    $v_{{\rm SFR},\phi}$ & 0.12 & 0.69 & -0.03 \\
    $v_{{\rm SFR},z}$ & 0.02 & -0.03 & 0.72 \\
  \end{tabular} \caption {
    Correlation matrix for the components of the peculiar motion of
    the Sun and the typical peculiar motion of the HMSFRs. The values
    given are, for example, $\mathrm{corr}(U_\odot,v_{{\rm SFR},R})$
    (eq.\ref{eq:corr}).  The
    strongest correlations are between components that, were the
    HMSFRs at the Sun's position, would have identical effects on the
    expected heliocentric velocity of the HMSFRs. Because of the sign
    conventions of the $U_\odot,V_\odot,W_\odot$ system,
    $\mathrm{corr}(U_\odot,v_{{\rm SFR},R})$ is negative (because they
    correspond to motions in opposite directions), while
    $\mathrm{corr}(V_\odot,v_{{\rm SFR},\phi})$ and
    $\mathrm{corr}(W_\odot,v_{{\rm SFR},z})$ are positive (because
    they correspond to motion in the same direction).
  }\label{tab:sun_maser}
\end{table}

The strong correlations that Table~\ref{tab:sun_maser} show have been
seen in previous studies \citep{PJMJJB10:Masers,Reea09}, and can be
easily understood: Consider the positions of the HMSFRs in
Galactocentric polar coordinates $(R,\phi,z)$, where $\phi_\odot = 0$.
Any HMSFR at $\phi=0$ will have no change in expected
\emph{heliocentric} velocity if both $\vsol$ and $\vsfr$ change by the
same amount. For $\phi\neq0$ there will be some difference, increasing
as $\phi$ goes further from $0$. The majority of the observed sources
are at $\phi$ relatively close to zero, which causes this correlation. 

If we remove the \cite{SBD10} constraint on the Sun's peculiar velocity, then the MCMC chain covers a broad range of values for the Sun's peculiar motion (especially $V_\odot$), with the peculiar motion of the HMSFRs adjusting in exactly the way one would expect -- if $V_\odot$ increases by $15\kms$, $v_{{\rm SFR},\phi}$ must increase by ${\sim}15\kms$ to keep roughly the same relative motion. This means that numerical experiments without the \citeauthor{SBD10} prior (or something like it) are not useful for constraining the parameters. However they do give us an insight into the implications of alternative estimates of $\vsol$ in the
literature. 

\cite{Sc12} suggested that \cite{SBD10} may have
underestimated the value of $U_\odot$, and that it may be
${\sim}14\kms$. This would imply that the HMSFRs have a typical velocity
radially inwards of ${\sim}6\kms$. 

In a study of the kinematics of stars from a large
range of Galactocentric radii observed by APOGEE \citep{APOGEE},
\cite{Boea12} found $V_\odot=(26 \pm 3)\kms$. They suggested
that this could be consistent with local measurements (based on stars
observed in the Solar neighbourhood) like that of \cite{SBD10} if
there was an offset between what they called the Rotational Standard
of Rest (RSR), which is the circular velocity at $R_0$ in an
axisymmetric approximation of the true potential (i.e. what we find in
this study), and the velocity of a closed orbit, and therefore a
theoretical $0\kms$ dispersion population, in the Solar neighbourhood
\citep[which is what is found by local measurements
like that of][]{SBD10}. This would reflect large scale non-axisymmetry of the potential. From our results we can infer that this would require
that a typical HMSFR leads circular rotation by ${\sim}11\kms$.

\cite{PJMJJB10:Masers} argued that large departures from circular
rotation (larger than ${\sim}7\kms$) were implausible based on the known
perturbations in the gas velocities due to non-axisymmetry in the
potential and the low velocity dispersion of the youngest stars
observed in the Solar neighbourhood. These arguments still hold, and
suggest that both the \cite{Sc12} and
\cite{Boea12} results should be treated with some scepticism. It is
worth noting that the results found by \cite{Boea12} included a
radial velocity dispersion that is nearly constant (or even
increasing) with radius, where there are good physical reasons to
expect it to fall with increasing radius, as well as observational
evidence \citep{LeFr89}. \cite{Shea14} noted that the use of Gaussian
models of the kind used by \cite{Boea12} can lead to this
behaviour in the model velocity dispersion where other models do not,
because the Gaussian model is unable to properly represent the
skewness of the $v_\phi$ distribution. This effect on velocity
dispersion tends to increase the model value for $V_\odot$, because of
the effect on asymmetric drift. This may explain why the
\cite{Boea12} result differs so significantly from that found here or
in the Solar neighbourhood studies. It is worth noting that \cite{Boea15} found a similar result to that of \cite{Boea12} using a different technique to analyse the kinematics of red clump stars observed in the mid-plane of the Milky Way. They minimized the `large scale power' in the velocity field, having subtracted an axisymmetric velocity field. The axisymmetric velocity field subtracted was that found by \cite{Boea12}, but they did also experiment with a velocity distribution that had a velocity dispersion that fell with radius, finding similar results. It is not as obvious why the results of this study might differ so substantially from our results. 

\begin{figure}
  \centerline{\resizebox{\hsize}{!}{\includegraphics[angle=270]{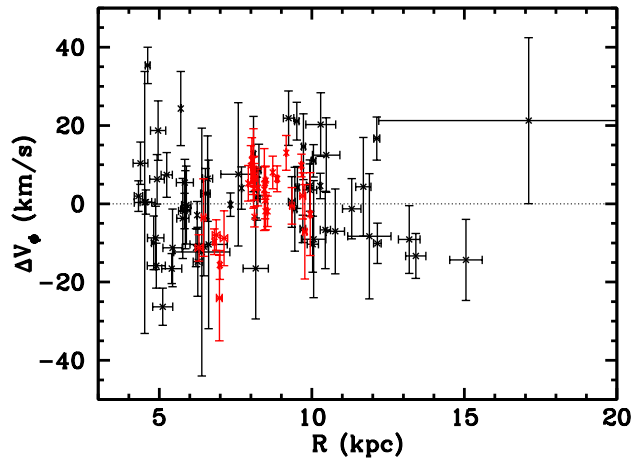}}}
  \caption{Positions of the maser sources in Galactocentric radius, plotted against offset from circular velocity in the best potential. Error bars are found from a Monte Carlo sample within the uncertainties on the observations. The points in red are those where the quoted parallax is $\varpi<0.5$, which are excluded from some experiments in Section~\ref{sec:solarv}.
\label{fig:Masers}
}
\end{figure}

We have investigated whether the difference between this result and that of \cite{Boea12} is dependent on data from the maser sources closest to the Sun. These observations have the smallest relative uncertainty, so they might be expected to carry great statistical weight. These closer observations would also suffer a very similar offset between the RSR and the velocity of closed orbits (the latter being what we expect the HMSFRs to follow, to a reasonable approximation). To check this, we have fit our models to data sets that only include masers with quoted parallaxes $\varpi_{\rm obs}<0.5\mas$ (i.e., quoted distances more than $2\kpc$ from the Sun, leaving only 62 masers). We find very similar results for the values of $V_\odot$ and $v_{{\rm SFR},\phi}$ -- if anything the derived lag of the masers sources slightly increases to ${\sim}2.5\kms$ in the latter case. 

As a final test, we fix $\vv_{\rm SFR}=0$, and can then remove the \cite{SBD10} prior on $\vv_\odot$. In this case, again using only the 62 masers with $\varpi_{\rm obs}<0.5\mas$,  we derive almost identical results for $V_\odot$ ($13.9\pm1.0\kms$) and $R_0$ ($8.21\pm0.10\kpc$). $v_0$ increases by slightly over over $1\kms$ to $(234.4 \pm 3.1)\kms$, and the spread in SFR velocities, $\Delta_v$, increases slightly to $(8.1 \pm 0.9)\kms$. We conclude that our results regarding $V_\odot$ are independent of the observations of masers in the vicinity of the Sun.

In Figure~\ref{fig:Masers} we give the distribution of maser sources in Galactocentric radius, and the offset of their azimuthal velocity from that of a circular orbit in our best fitting potential. It is clear that the sources sample a large range of radii, reaching far from the Solar neighbourhood.

\section{Alternative models} \label{sec:alternative}
\begin{table*}
  \begin{tabular}{cccccccc}

    Difference from  &  $R_{\d,\mathrm{thin}} $  & $R_0$ & $v_0 $ & $\rho_{\mathrm{h},\odot} $ & $\rh $ & $M_\ast $ & $M_v$ \\
     main model &  $(\kpc) $  & $(\kpc)$ & $(\kms)$ & $(\msun\pc^{-3})$ & $(\kpc)$ & $ (10^9\msun)$ & $(10^{12}\msun)$ \\\hline
\rule[-1.2ex]{0pt}{0pt}  Main model: $\gamma= 1.00$  & $ 2.53 \pm 0.14 $ & $ 8.20 \pm 0.09 $ & $ 232.8 \pm 3.0 $ & $ 0.0101 \pm 0.0010 $ & $ 18.6 _{- 4.4 }^{+ 5.3 }$ & $ 54.3 \pm 5.7 $ & $ 1.32 \pm 0.29 $ \\ \hline
\rule[-1.2ex]{0pt}{0pt}  $\gamma$ free ($\gamma=0.79\pm0.32) $ & $ 2.51 \pm 0.15 $ & $ 8.20 \pm 0.09 $ & $ 232.5 \pm 3.0 $ & $ 0.0098 \pm 0.0009 $ & $ 15.4 _{- 3.8 }^{+ 8.0 }$ & $ 56.6 \pm 6.2 $ & $ 1.34 \pm 0.28 $ \\
\rule[-1.2ex]{0pt}{0pt} $\gamma= 0$  & $ 2.36 \pm 0.13 $ & $ 8.21 \pm 0.10 $ & $ 233.2 \pm 3.0 $ & $ 0.0103 \pm 0.0009 $ & $ 7.7 _{- 0.9 }^{+ 1.4 }$ & $ 57.7 \pm 4.5 $ & $ 1.11 \pm 0.19 $ \\
\rule[-1.2ex]{0pt}{0pt} $\gamma= 0.25$  & $ 2.40 \pm 0.13 $ & $ 8.21 \pm 0.10 $ & $ 232.9 \pm 3.1 $ & $ 0.0100 \pm 0.0010 $ & $ 9.6 _{- 1.8 }^{+ 2.6 }$ & $ 58.4 \pm 5.4 $ & $ 1.21 \pm 0.24 $ \\
\rule[-1.2ex]{0pt}{0pt} $\gamma= 0.5$ & $ 2.42 \pm 0.13 $ & $ 8.20 \pm 0.09 $ & $ 232.7 \pm 3.0 $ & $ 0.0101 \pm 0.0009 $ & $ 11.7 _{- 1.9 }^{+ 2.6 }$ & $ 57.5 \pm 5.1 $ & $ 1.24 \pm 0.23 $ \\
\rule[-1.2ex]{0pt}{0pt} $\gamma= 0.75$ & $ 2.46 \pm 0.13 $ & $ 8.21 \pm 0.09 $ & $ 232.9 \pm 3.0 $ & $ 0.0102 \pm 0.0009 $ & $ 13.8 _{- 2.5 }^{+ 3.1 }$ & $ 55.4 \pm 5.1 $ & $ 1.24 \pm 0.23 $ \\
\rule[-1.2ex]{0pt}{0pt} $\gamma= 1.25$  & $ 2.62 \pm 0.15 $ & $ 8.20 \pm 0.09 $ & $ 232.4 \pm 3.0 $ & $ 0.0099 \pm 0.0010 $ & $ 27.2 _{- 7.2 }^{+ 9.5 }$ & $ 52.8 \pm 5.8 $ & $ 1.47 \pm 0.38 $ \\
\rule[-1.2ex]{0pt}{0pt} $\gamma= 1.5$  & $ 2.82 \pm 0.17 $ & $ 8.19 \pm 0.09 $ & $ 232.4 \pm 2.8 $ & $ 0.0098 \pm 0.0009 $ & $ 46.1 _{- 11.7 }^{+ 13.8 }$ & $ 50.4 \pm 5.3 $ & $ 1.59 \pm 0.36 $ \\ \hline
\rule[-1.2ex]{0pt}{0pt} $\epsilon=-0.1$ & $ 2.45 \pm 0.13 $ & $ 8.19 \pm 0.09 $ & $ 232.2 \pm 2.9 $ & $ 0.0104 \pm 0.0009 $ & $ 19.0 _{- 3.5 }^{+ 4.1 }$ & $ 52.1 \pm 4.9 $ & $ 1.39 \pm 0.23 $ \\
\rule[-1.2ex]{0pt}{0pt} $\epsilon=+0.1$  & $ 2.61 \pm 0.13 $ & $ 8.22 \pm 0.09 $ & $ 233.3 \pm 3.0 $ & $ 0.0093 \pm 0.0010 $ & $ 20.8 _{- 4.8 }^{+ 7.0 }$ & $ 59.4 \pm 6.2 $ & $ 1.41 \pm 0.37 $ \\ \hline
\rule[-1.2ex]{0pt}{0pt} weak bulge prior &  $ 2.46 \pm 0.16 $ & $ 8.20 \pm 0.09 $ & $ 232.7 \pm 3.0 $ & $ 0.0101 \pm 0.0009 $ & $ 18.8 _{- 3.8 }^{+ 5.1 }$ & $ 54.3 \pm 5.3 $ & $ 1.35 \pm 0.27 $ \\
\rule[-1.2ex]{0pt}{0pt} weak $R_0$ prior & $ 2.38 \pm 0.15 $ & $ 7.97 \pm 0.15 $ & $ 226.8 \pm 4.2 $ & $ 0.0098 \pm 0.0010 $ & $ 20.8 _{- 5.5 }^{+ 6.8 }$ & $ 52.0 \pm 5.5 $ & $ 1.39 \pm 0.35 $ \\ \hline
\rule[-1.2ex]{0pt}{0pt} Gas discs $\rho\times0.7$ & $ 2.57 \pm 0.14 $ & $ 8.20 \pm 0.09 $ & $ 232.6 \pm 2.9 $ & $ 0.0100 \pm 0.0009 $ & $ 19.9 _{- 4.2 }^{+ 4.9 }$ & $ 53.8 \pm 5.4 $ & $ 1.39 \pm 0.27 $ \\
\rule[-1.2ex]{0pt}{0pt} Gas disc  $\rho\times1.3$ &  $ 2.45 \pm 0.14 $ & $ 8.20 \pm 0.09 $ & $ 232.6 \pm 3.0 $ & $ 0.0100 \pm 0.0009 $ & $ 18.9 _{- 4.3 }^{+ 5.4 }$ & $ 57.1 \pm 5.4 $ & $ 1.34 \pm 0.28 $ \\
  \end{tabular} \caption{
    Properties of alternate mass models from our main models (we also include results from our main models for ease of comparison). These models are described in detail in Section~\ref{sec:alternative}. The properties shown are those that we consider most relevant to our discussion.
  }\label{tab:altmodels}
\end{table*}

\subsection{Varying inner halo density profile} \label{sec:vargamma}

\begin{figure}
  \centerline{\resizebox{\hsize}{!}{\includegraphics{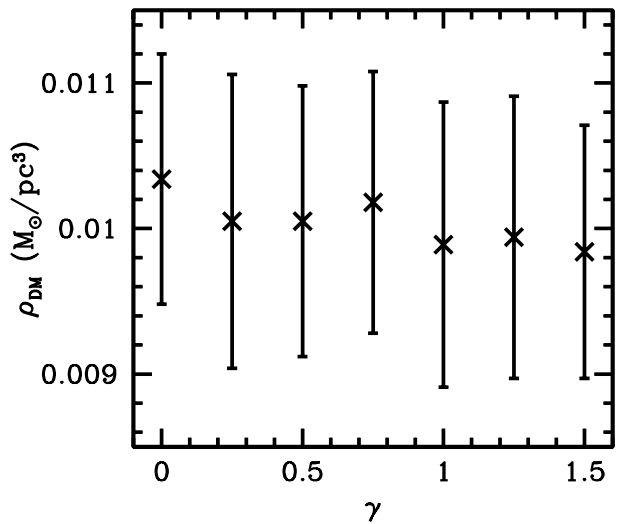}}}
  \vspace{-30pt}
  \caption{Local dark-matter density determined from models with differing inner density slopes for the dark-matter component ($\rho\propto r^{-\gamma}$ for $r\ll \rh$). The trend is for slightly lower dark-matter density values for steeper inner density slopes, but the trend is much smaller than the individual error bars.
\label{fig:RhoDM}
}
\end{figure}
\begin{figure}
  \centerline{\resizebox{\hsize}{!}{\includegraphics{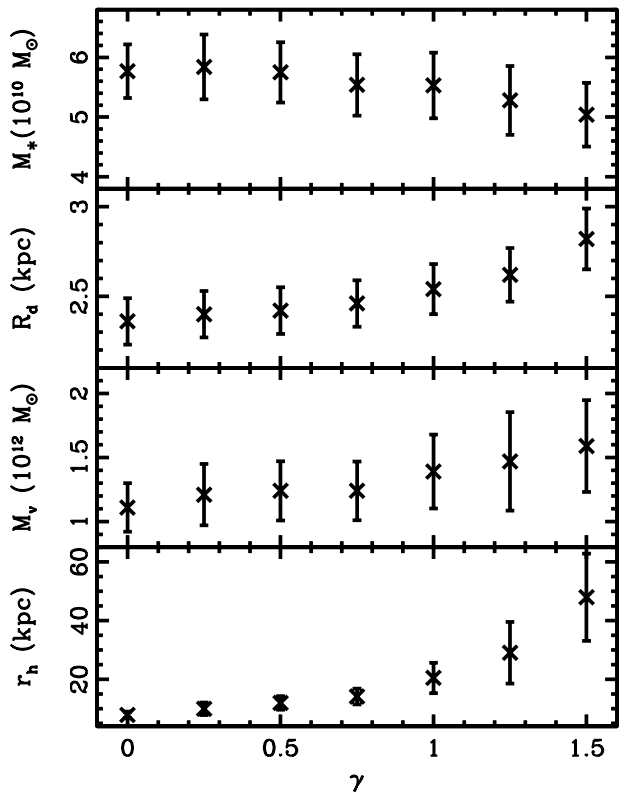}}}
  \caption{Properties of the model that strongly vary with inner density slope of the dark-matter component ($\rho\propto r^{-\gamma}$ for $r\ll \rh$). The top two properties are for the stellar component -- the total stellar mass $M_\ast$ and the thin-disc scale length $R_\d$, while the lower ones are for the whole Galaxy (the virial mass, $M_v$) and the dark-matter component only (scale radius $\rh$).
\label{fig:GammaVariation}
}
\end{figure}
We have explored the effect of varying the inner slope of the dark-matter density profile ($\gamma$ in eq.~\ref{eq:halo}). The density of the dark-matter halo goes as $r^{-\gamma}$ at $r \ll r_\mathrm{h}$. This is a simple approach to exploring the question of whether the Milky Way's dark-matter halo is cusped (like the NFW profile) or not. As noted in Section~\ref{sec:halo}, there are good reasons to believe that the halo profile has been changed from that expected in pure dark matter simulations, because baryonic processes will inevitably have had an effect. In reality there is no obvious reason for any effect that might have altered the dark-matter profile  -- be they baryonic processes or due to the unexpected nature of the dark-matter particle -- to occur on the same physical scale as the halo's scale radius $\rh$. The true density profile may be poorly characterised by a two-power-law equation like eq.~\ref{eq:halo}. Nonetheless, this numerical experiment does provide us with useful insights.

If we set $\gamma$ to be a free parameter (with a flat prior), we find $\gamma=0.79\pm0.32$. 
This is a rather weak constraint, which is not too surprising since determining the value of $\gamma$ depends sensitively on the details of the inner Galaxy (where we have the worst constraints), and to some extent on the outer Galaxy (where our constraints are also limited).

Perhaps more remarkable is the fact that the derived properties and uncertainties of these models are extremely similar to those found when we fix $\gamma=1$. The thin-disc scale length $R_{\d,\mathrm{thin}} = (2.51\pm0.15)\kpc$, which differs from that found where $\gamma=1$ by only ${\sim}20\pc$ (with ${\sim}10\pc$ greater uncertainty). The derived value of $R_0$ agrees to within $10\pc$ and the various derived velocities ($v_0$ and the peculiar velocities of the Sun and maser sources) agree to within $0.2\kms$. The derived total stellar mass is $(56.6\pm6.2)\times10^9\msun$, which is slightly higher than the value found with $\gamma=1$, but well within the quoted uncertainty. The total virial mass is identical to the quoted precision.

The local dark-matter density is barely affected by setting $\gamma$ free, with the local density $0.0099\pm0.0009\msun\pc^{-3}$ (a ${\sim}1$ per cent change in the expectation value for the density, compared to a ${\sim}10$ per cent statistical uncertainty). This is rather surprising as one might expect the local dark-matter density to depend rather sensitively on the dark-matter density profile for $r<\rh$. This merits some further examination. We have therefore investigated models with $\gamma$ set at fixed values in the range $0\leq\gamma\leq1.5$.

In Fig.~\ref{fig:RhoDM} we show the expectation values and uncertainties of the local dark-matter density as a function of $\gamma$. We see that the change in expectation value of $\rho_{\mathrm{h},\odot}$ is very small -- it \emph{decreases} by a few percent as we go from a completely cored profile ($\gamma=0$) to a very steeply cusped one. Indeed most of the properties we measure of the mass models (including $v_0$) are only very slightly affected by the change of $\gamma$. Clearly what we have is a tight constraint on the \emph{local} properties of the Milky Way, rather than at any other point in the Galaxy. If the constraint on the dark-matter density was, instead, tightest at a radius ${\sim}1\kpc$ less than $R_0$, that would correspond to a $20$ per cent change in $\rho_{\mathrm{h},\odot}$ as we go from $\gamma=0$ to $\gamma=1.5$.

In Fig.~\ref{fig:GammaVariation} we show the properties of the models which \emph{do} significantly change with varying $\gamma$. As $\gamma$ increases, the halo becomes more centrally concentrated so the scale length of the disc increases, making the disc less centrally concentrated, to compensate and leave the rotation curve essentially the same. The total disc mass decreases, for a similar reason. Meanwhile the halo scale radius decreases when $\gamma$ decreases, to the point where $r_\mathrm{h}<R_0$ for $\gamma=0$. This has the effect that the slope of the halo density profile $\d \log \rho/\d\log r$ is always ${\sim}-1.5$ at $R_0$. The knock-on effect is that the virial mass of the halo increases with increasing $\gamma$.

\subsection{Perturbing the disc} \label{sec:spiral}
We now investigate whether small perturbations to the assumed exponential density profile (of the order of those expected due to spiral arms) have any significant effect on these results. If this is the case, it would imply that our assumed constraint on the density profile of the stellar discs was too strict. We do not include any non-axisymmetric component to the potential. 

We consider discs with surface densities of the form
\begin{equation} \label{eq:wobblydisc}
  \Sigma_\mathrm{d}(R)=\Sigma_{0}\;\textrm{exp}\left(-\frac{R}{R_\mathrm{d}} + \epsilon \cos \frac{\pi R}{R_\mathrm{d}} \right),
\end{equation}
where we set a perturbation amplitude $\epsilon=\pm0.1$, which provides a perturbation with a maximum amplitude of ${\sim}10$ per cent of the local surface density at a given radius, and has a scale length of $2 R_\d$. We make no claim that this represents the true surface density, but rather that it acts as an illustrative case. 

Results for models with these perturbations are shown in Table~\ref{tab:altmodels}. Introducing this perturbation has almost no effect on the derived values of $R_0$ or $v_0$. It alters the derived scale length of the thin disc by much less than $10$ per cent, which is well within what one would naively expect. The derived stellar mass does change by ${\sim}10$  per cent (of a similar order to the derived statistical uncertainty), while the virial mass barely changes. The derived local dark-matter density is also affected by the perturbation (surprisingly, more than by any change in $\gamma$ we considered), but this is still somewhat smaller than the statistical uncertainty.

DB98 used a similar form of perturbation to that given in eq.~\ref{eq:wobblydisc}, but had a rather longer scale length for the perturbation, which was of the form $\epsilon \cos R/R_\mathrm{d}$ (i.e. without the factor of $\pi$). They noted that this produced large changes to the rotation curve in the outer parts of the Milky Way. This perturbation goes from peak to trough over a distance of $\pi \times R_\mathrm{d}\sim R_0$. Its main effect is therefore to either increase or decrease the effective scale length of the disc over the radial range that contains most of its mass. Since DB98 held the scale length of the disc fixed for each model (and showed that varying it has important knock-on effects on the whole model), we suspect that the effect they describe is primarily due to the change in the effective scale length, rather than illustrative of an overconfidence introduced by assuming an exponential surface density profile. In our study we have not found that introducing a perturbation of this form affect the results at large radii (illustrated by the virial mass being nearly unchanged in models with $\epsilon=\pm0.1$).

In his master's thesis \cite{Zi16} found that the inclusion of a non-axisymmetric element to the velocity field for the maser sources and terminal velocity curve (to represent the effect of spiral structure) had a rather small impact on the parameters of a simple mass model of the Galaxy. It altered the derived local dark-matter density by a couple of per cent, and had a similar effect on the mass contained within $50\kpc$. While the effect of non-axisymmetric structure is clearly important for some studies of the potential of the Milky Way even outside the bulge \citep[see e.g.][]{Gaea12}, for our purposes here it is reasonable to ignore it.


\subsection{An uncertain bulge} \label{sec:weakBulge}
The prior that we take for the bulge is relaxed compared to the quoted uncertainty from \cite{BiGe02}, but as they make prior assumptions, it could be argued that this is a narrow range of possibilities. \cite{Poea15M2M}  found a wider range of \emph{stellar} masses within the region they studied (though the total mass was found with a very small uncertainty, sec Section~\ref{sec:others}). We have therefore investigated the effects of weakening our bulge prior, such that it had an uncertainty of $\pm 30$ per cent on mass.

\begin{figure}
  \centerline{\resizebox{0.7\hsize}{!}{\includegraphics[angle=270]{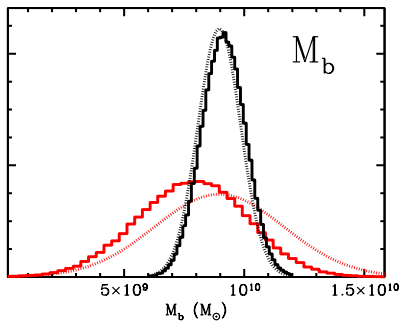}}}
  \caption{Pdf of bulge mass for our normal (black) and relaxed (red) priors on bulge mass (10 and 30 per cent uncertainty respectively). In both cases the prior is shown as a dotted line and the derived pdf as a solid histogram. The derived pdf is, in both cases, reasonably close to the prior, indicating that the additional data does not provide significant constraints on the bulge.
\label{fig:Mbul}
}
\end{figure}

As we have noted before, the maser data and terminal velocity curve does not provide much of a constraint on the details of the bulge region, because we have excluded these regions to avoid false conclusions caused by our assumption of axisymmetry. This means that weakening our bulge prior ensures that we will find a far greater range of possible bulge masses. In Figure~\ref{fig:Mbul} we show the derived pdfs of the bulge mass for both the original prior (10 per cent uncertainty) and the weakened prior (30 per cent). In the latter case, the mass we find is $ (7.83 \pm 2.30) \times10^9\msun$. This mean value is lower than the value used as a prior (suggesting that other data favours a slightly lower mass bulge), but the spread of values comfortably includes the values associated with the tighter prior. 

The properties of the Galaxy models found with this weak prior are shown in Table~\ref{tab:altmodels}. Almost all of the properties are essentially unchanged (as one would hope), but the scale radius of the disc is decreased from the main model value by $\sim0.07\kpc$ (compared to an uncertainty of $0.16\kpc$, up from $0.14\kpc$ in the main model). This is related to the lower average bulge mass -- we can see in Table~\ref{tab:coo_par} or Figure~\ref{fig:Corner} that the two parameters are correlated.

We conclude that even if we substantially weaken the prior on our bulge component, the properties found for the rest of the Galaxy are not substantially affected.

\subsection{Altered gas discs} \label{sec:hiloGas}

The gas content of the Milky Way remains deeply uncertain, as discussed in Section~\ref{sec:gasdisc}. The gas disc models used (as described in Table~\ref{tab:gas}) are efforts in good faith to capture the current understanding in a convenient form, but we have not attempted to fit them to the data, because we do not have the statistical leverage to do so. This does, however, introduce a systematic uncertainty that we have not accounted for.

In an effort to get some idea of how important this uncertainty is, we have investigated the effect of increasing or decreasing the total mass of the gas discs by 30 per cent (i.e. $\sim \pm 3.5\times10^9\msun$). This is approximately the level of spread seen when comparing different investigations of the \HI\ disc (see e.g., KD08), which is the main component. We achieve this by simply changing the density at any point by the quoted factor (i.e. changing $\Sigma_0$ in eq.~\ref{eq:gasdisc}).

The results are given in Table~\ref{tab:altmodels}, and the consequences are relatively minor. If we decrease the gas mass, the disc scale length increases for the same reasons discussed in Section~\ref{sec:Main}, and when we increase the gas mass, the disc scale length decreases. There is a related effect that the total stellar mass increases when the gas mass increases (and they also decrease together), but by less than the change in gas mass (or the quoted uncertainties). 

\subsection{Weaker $R_0$ prior} \label{sec:weakR}
The prior that we take on the distance to the Galactic centre from \cite{Chea15} is derived from observations of objects near the Galactic centre itself (the nuclear stellar cluster, and stars directly orbiting Sgr A*). These provide rather more direct estimates of this distance than techniques like those used in this study, which rely on the Galactic centre being the centre of (circular) rotation for the Galaxy at large. That is why we have taken the (relatively narrow) prior from \cite{Chea15}, despite the evidence that the maser observations drive us towards lower values of $R_0$ (at 1$\sigma$ below our prior).

Nonetheless, we also consider the case with a weaker prior on $R_0$, specifically the prior that we took in Paper I, which is simply derived from the measurements of the orbits of stars around Sgr A* by \cite{Giea09}, and is $R_0=8.33 \pm 0.35 \kpc$. The results are shown in Table~\ref{tab:altmodels}, and show significant differences from the results with the more restrictive prior. The thin disc scale length decreases to $(2.38\pm0.15)\kpc$, which is roughly 1$\sigma$ below the value for our main model, while $R_0$ and $v_0$ decrease more substantially, to $(7.97\pm0.15)\kpc$ and $(226.8\pm4.2)\kms$ respectively. The other main properties of the Galaxy are effected by less than the quoted uncertainties.

This serves to emphasise the importance of constraints on $R_0$ when it comes to determining the circular velocity of the Milky Way. Because the \cite{Chea15} study provides constraints based on well understood observations near the Galactic Centre, we chose to retain this prior when constructing our main models, but the tension between the value of $R_0$ they derive and the value that the rest of our constraints drive us towards is worthy of further study.


\section{Comparison to other studies} \label{sec:others}

Recent reviews by \cite{BHGe16} and \cite{Re14} provide excellent summaries of the literature related to the topics discussed in this paper. From an analysis of literature values of $R_0$, \cite{BHGe16} adopt a best estimate of $(8.20\pm0.1)\kpc$, very much in keeping with the value found here. We note with interest, however, that a more recent analysis of the astrometry of faint stars orbiting Sgr A*, combining two different types of imaging over a time baseline of two decades finds a lower value of $R_0=(7.86\pm0.14\pm0.04)\kpc$ \cite[statistical and systematic uncertainties respectively,][]{Boea16}. This is more similar to the results we find with a weaker prior, and suggests that the value of $R_0$ is still not fully settled.

\cite{BHGe16} also discuss the literature regarding the disc scale lengths, noting that estimates range from $1.8$ to $6.0\kpc$. They conclude from the 15 `main papers' on the topic (it is not explicitly stated which papers these are) that the best estimate for the thin-disc scale length is $(2.6\pm0.5)\kpc$, very similar to the value from \cite{Juea08} that we take as a prior. Our study provides a rather tighter constraint on this value, but we note that it also demonstrates that estimates of this value that are based on dynamics depend strongly on assumptions made regarding the other components of the Galaxy, such as the mass and scale lengths of the gas discs, and the density profile of the dark-matter halo.

\cite{Re14} gives a compilation of estimates of the local dark-matter density (his fig. 2 and table 4). The value we find ($0.0100\pm0.0010\msun\pc^{-3}$) is larger than many of the local measures in the compilation \citep[though not all, e.g.,][when no strong prior on the baryonic component is invoked]{Gaea12}. It is quite typical of values found when looking at the global structure of the Galactic potential (and assuming spherical symmetry, see Section~\ref{sec:streams}), including Paper I, \cite{CaUl10}, and \cite{NeSa13}.

\cite{Poea15M2M} used Made-To-Measure modelling to determine the mass within a box around the Galactic centre of a volume $(\pm2.2\pm1.4\pm1.2)\kpc$, determining that it is $(1.84\pm0.07)\times10^{10}\msun$ \citep[and the closely related study by][found a very similar value -- $(1.85\pm0.05)\times10^{10}\msun$]{Poea16}. They note the difficulty of comparing different studies of the bulge because different studies probe different regions. They point out that \cite*{BiEnGe03} determined the circular velocity at $2.2\kpc$ to be $190\kms$, and equate this (under the assumption of spherical symmetry) to a mass of $1.85\times10^{10}\msun$. We find a circular velocity at $2.2\kpc$ of $(198\pm9)\kms$, which would correspond (again, assuming spherical symmetry) to a mass of $\sim(2.01\pm0.17)\times10^{10}\msun$. Our result is clearly comparable to these other studies, and given that the \cite{Poea15M2M} study looks at only a part of this region (though certainly the densest part, and it also includes a small volume outside a spherical shell of radius $2.2\kpc$), it is reassuring that our result corresponds to a slightly higher mass than theirs. Again, we must emphasise that since our study is axisymmetric, it cannot accurately describe the bulge region, but the reasonable agreement between our study and that of \cite{Poea15M2M} is a useful sanity check.

\subsection{The outer Galaxy} \label{sec:outer}
\cite{Waea10} used a sample of 26 satellite galaxies (including six with proper motions) to find the mass of the Milky Way within $300\kpc$, and found that the answer depended sensitively on the assumed velocity anisotropy of the satellites, with a mass estimate that was $(1.4\pm0.3) \times 10^{12}\msun$ assuming isotropic velocities, but which could plausibly lie anywhere between $1.2$ and $2.7\times 10^{12}\msun$ when anisotropy is taken into account. Our estimate, which is $(1.6\pm0.3)\times10^{12}\msun$, fits comfortably in this range.

\cite{Deea12} and \cite{Xuea08} used samples of blue horizontal branch stars to determine the mass contained within $50$ and $60\kpc$ (respectively). In the latter case, cosmological simulations were used to provide a prior on the velocity anisotropy of the population. \cite{FeSc13} raise important concerns regarding the selection of stars for these studies\footnote{\cite{FeSc13} focussed on different studies by the same lead authors, but the selection criteria used are similar or identical, so the points made are entirely relevant.}. The cuts in $\log g$ and $g$-band magnitude used by \cite{Deea12} lead to a very high risk of contamination by disc stars (\citeauthor{FeSc13} put this contamination at ${\sim}25$ per cent of the sample), which will naturally have significant effects on the results. The concerns raised regarding the \citeauthor{Xuea08} sample are more subtle: stars with a high value of a specific spectroscopic `steepness parameter' $c_\gamma$ are found to have a systematically different line-of-sight velocity to those with other values. This suggests that there is either a problem with the spectroscopic pipeline or that these stars are part of a stream-like component.

\cite{Kaea14} use similar techniques to this paper (and Paper I) to determine a best fitting mass model for the Milky Way as a whole, taking into account many constraints. They focussed on the Jeans modelling of halo giant and blue horizontal branch stars. When using the same definition of virial mass that we employ in this study, they find a quoted $M_v =0.72^{+0.2}_{-0.13}\times10^{12}\msun$, which is significantly smaller than the value we find in this study. However it must be noted that this value is heavily dependent on the value of $R_0$, which they take to be $8.5\kpc$. When they perform the same analysis with $R_0=8.0\kpc$ the estimated virial mass increases by $50$ per cent. We therefore do not interpret their result as being in serious tension with ours.

\cite{Piea14Escape} analysed the velocity distribution of a small sample of high-velocity stars from the RAVE survey, to determine the local escape speed \citep[this follows the similar work by][]{Smea07}. This provides an estimate of the mass of the Milky Way, if one assumes a profile for the dark-matter halo. They found a Milky Way virial mass \citep[using the same definition as][]{Moea13} of $m_v = 1.6^{+0.5}_{-0.4}\times10^{12}\msun$ assuming an NFW halo, and $1.4^{+0.4}_{-0.3}\times10^{12}\msun$ with a halo that had adiabatically contracted. Either value is entirely consistent with our results.

The prior that we take from \cite{WiEv99} is based on a study of just 27 objects with known distances and radial velocities, of which only six also had measured proper motions. Nonetheless, we feel comfortable using the upper limit on $M_{50}$ that they derive because (1) the careful uncertainty analysis they performed (which led to large quoted uncertainties on many properties of the Milky Way) instills confidence in the one strong limit they do set, and (2) there are (to our knowledge) no direct studies of the Milky Way which demand a higher mass within $50\kpc$ than the upper limit we set.

\subsection{Dynamical modelling of the disc} \label{sec:dynmod}

\begin{figure}
  \centerline{\resizebox{\hsize}{!}{\includegraphics{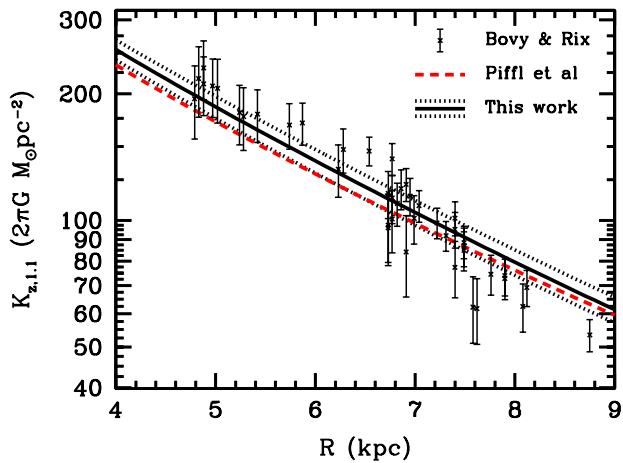}}}
  \vspace{-40pt}
  \caption{
The vertical force $1.1\kpc$ from the Galactic plane as a function of radius for our main models, with $1\sigma$ uncertainties (\emph{solid and dashed lines, respectively}), for the best fitting model of \protect\cite[\emph{dashed red line}]{Piea14}, and from the fits to individual mono-abundance populations of \protect\cite[\emph{points with error bars}]{BoRi13}.
\label{fig:Kz}
}
\end{figure}

In recent times a great deal of focus has gone into the determination of the Milky Way potential by fitting of the observed velocities of disc stars to distribution functions expressed in terms of action coordinates. Despite the large computational demands of this approach, it is substantially more practical for analysing Milky Way data than approaches based on orbit libraries \citep[see e.g.,][]{PJMJJB13}. Two examples of this are \cite{Piea14}, who analysed data from the RAVE survey by comparing velocity histograms from models with those derived from the data (and used similar constraints from maser sources to those used in this study), and \cite{BoRi13} who fit quasi-isothermal distribution functions \citep{JJBPJM11} to `mono-abundance populations' (groups of stars in a small range of $[\mathrm{Fe}/\mathrm{H}]$ and $\alphaFe$) from the SEGUE survey \citep{SEGUE}.

The main result from \cite{Piea14} was that, for a spherical halo, the local dark-matter density is $0.0126\msun\pc^{-3}$ with a systematic uncertainty of $15$ per cent. This is slightly higher than our value, but within ${\sim}1\sigma$ when the uncertainties on both quantities are considered. The $K_{z,1.1}(R)$ profile (vertical force $1.1\kpc$ from the plane as a function of radius) of their best fitting model is shown in Fig.~\ref{fig:Kz}, and is very similar to ours. 

Fig.~\ref{fig:Kz} also shows the values of $K_{z,1.1}(R)$ found by \cite{BoRi13} (scaled such that $R_0=8.2$). While the scatter of points around our model appears consistent with the error bars, there is a clear trend for the points at $R\lesssim7\kpc$ to lie above the line from our model and the points at $R\gtrsim7\kpc$ to lie below it. The values of $K_{z,1.1}(R)$ found by \cite{BoRi13} are from individual `mono-abundance populations', treated independently, and as coming from a single quasi-isothermal distribution function with an exponential density profile. The values given by \cite{BoRi13} and plotted in Fig.~\ref{fig:Kz} are from the radius where a given mono-abundance population gives the tightest constraint, but each population provides constraints over a wide range of radii. The constraints provided by these different populations are not mutually consistent. The points in Fig.~\ref{fig:Kz} therefore give the statistical uncertainty for each population at the radius where it is smallest, while there is clearly a systematic error (otherwise the different constraints would be consistent), which may be due to the assumption that each mono-abundance population is well described by a quasi-isothermal distribution function. This may explain the difference between the points in Fig.~\ref{fig:Kz} and the two sets of lines. 

\cite{McG16} used a mass discrepancy-acceleration relation (MDAR) to relate the terminal velocity curve to a disc surface density profile. The vertical force at $1.1\kpc$ found was broadly similar to that found by \cite{BoRi13}  -- see \citeauthor{McG16}'s fig. 14 -- but with more `bumps and wiggles' and noticeably even further below the \cite{BoRi13} values for $5.5\kpc \lesssim R \lesssim 6.5\kpc$ than our model.

\subsection{Streams and the shape of the dark matter halo} \label{sec:streams}

Because of the large number of streams found in the Milky Way halo in recent years, it has become increasingly popular to attempt to constrain the potential of the Milky Way by fitting the properties of these observed streams to models. It is important to note that these streams are formed when satellites of the Milky Way are tidally disrupted, putting the stars that were in the satellite onto differing orbits. The \emph{difference} between the motion on these orbits is the primary influence on the observed structure of the stream \citep{EyJJB11}. 

\cite*{Giea14} find a very low mass for the Milky Way out to $100\kpc$ of $(4.1\pm0.4)\times 10^{11}\msun$ by modelling the Sagittarius stream. This is significantly lower than other estimates, \citep[see][Table 8]{BHGe16}. It is worth noting that the comparison between model and data was based entirely on the positions of the apogalactic points of the leading and trailing tails of the stream, and the precession angle between them. It is not clear whether the model can explain the other properties of the Sagittarius dwarf. The study of the Sagittarius stream by \cite{JoLaMa05} had best fitting models with masses in the range $(3.8-5.6)\times10^{11}$ within $50\kpc$, which is in keeping with our prior. We therefore treat this result with caution, and merely note that it differs substantially from the value we find for the mass inside $100\kpc$ of $(8.2\pm1.1)\pm10^{11}\msun$. If it is supported by further results we will have to re-evaluate our assumptions and find new models.

\cite{Kuea15} used the position of apparent overdensities in the tidal tails of Palomar 5 to determine the mass within its apogalactic radius. However, \cite{Ibea16} used deeper photometry of these tidal tails to demonstrate that these overdensities are very likely to be observational artefacts associated with inhomogeneities in the SDSS photometry used to produce the maps of Palomar 5 used by \citeauthor{Kuea15}, \citep[see also][]{Thea16}.

The flattening of the Milky Way's dark-matter halo remains deeply uncertain. Famously, models of the Sagittarius stream have been used to argue that the dark-matter halo is oblate \citep*[e.g.][]{JoLaMa05}, prolate \cite{He04} or triaxial \citep[though this is effectively oblate with the short axis in the plane of the Galaxy]{LaMa10}. Analyses of the smooth stellar halo as seen by the SDSS survey has yielded claims of constraints on the shape of the equipotential surfaces of the halo ranging from oblate with axis ratio $q_\Phi = 0.7\pm0.1$ \cite[corresponding to axis ratios in the density distribution of $0.4\pm0.1$,][]{Loea14} to prolate with axis ratio $1.5\lesssim q_\Phi \lesssim2$ \citep*{BoEvWi16}. Analysis of the GD-1 stream, meanwhile, has yielded flattening estimates $q_\Phi \sim0.9$ \citep*{BoBeEv15}. In the Solar neighbourhood, \cite{Re14} noted that the comparison of constraints from the rotation curve (which provides an estimate of the spherically averaged enclosed dark-matter content) and estimates of the local density provide a constraint on the halo shape which is consistent with spherical or slightly prolate. \cite{Piea14} argued that combining their results with those of \cite{Biea14} provided tentative indications of an axis ratio (in the isodensity contours) of $q\sim0.8$. \cite{Piea14} also noted that the axis ratio of the dark-matter halo has an important influence on the derived local density when interpreting dynamical models of the disc, with $\rho_{h,\odot}\propto q^{-0.89}$.

This is a deeply unsatisfying state of affairs, and we have followed numerous other authors in simply assuming that the dark-matter halo is spherically symmetric. This is an assumption that will need to be revisited to build a better model. Indeed the assumption of a constant axis ratio at all radii will need examination. The argument of \cite{Re14}, that comparing local and global constraints provides a route to solving this problem, indicates the role that approaches such as the one used in this study may have in the future. One must be careful, however, as this study has used a local constraint \citep[from][]{KuGi91}, so is not suitable for making an unbiased comparison between the two approaches.

\section{Conclusions} \label{sec:conc}

We have returned to the Bayesian methods laid out by Paper I, and used new data and an improved underlying model to find a new model of the mass distribution of the Milky Way. This new model includes components that represent the contribution of the cold gas discs near the Galactic plane, as well as thin and thick stellar discs, a bulge component and a dark-matter halo.
We have used an MCMC approach to determine the properties of the models and their statistical uncertainties. We have also determined a best fitting mass model, which provides a gravitational potential that can be used as a starting point in any dynamical modelling, and for determining the orbits of stars in the Milky Way. We emphasise that this is an axisymmetric model, so cannot reflect the strongly non-axisymmetric bar structure in the inner few kpc of the Milky Way.

We have explored models that differ from the standard NFW halo density profile by allowing the density slope in the inner halo ($\gamma$ in eq.~\ref{eq:halo}) to vary freely and fitting it to the data, or by holding it at fixed values that differ from that of the NFW profile. If this is left as a free parameter, it is only weakly constrained $\gamma=0.79\pm0.32$. We have found that the local dark matter density is surprisingly close to being independent of the assumed value of $\gamma$.

To explore possible systematic biases or overconfidence, we have explored models that include deviations from a purely exponential disc surface density. The effect on the virial mass is minimal, as is that to the derived values of $R_0$ or $v_0$, but a ${\sim}10$ per cent perturbation to the form of the surface density profile (which is at a level that is plausible) does produce a $\lesssim10$ per cent change in the derived local dark-matter density and the total stellar mass. We have explored other possible systematic uncertainties, and we would estimate that the  systematic uncertainties (neglecting flattening of the dark-matter halo) are generally of the same order as the statistical uncertainties.

The value of $R_0$ that we determine -- $(8.20\pm0.09)\kpc$ -- is around $1\sigma$ lower than the one we take as our prior \citep{Chea15}, and if we take a weaker prior then the value of $R_0$ is driven down still further, to ${\sim}8.0\kpc$. Since we know $v_0/R_0$ far better than we know $v_0$ independently of $R_0$, this remains an important uncertainty. The recent estimate of $R_0\sim7.9\kpc$ from a new reduction of observations showing the orbits of stars around Sgr A* \citep{Boea16} would appear to cast fresh uncertainty on this value when it seemed as if most estimates were tending towards ${\sim} 8.2-8.3\kpc$ \citep{BHGe16}.

We left the peculiar velocity of the Sun, $\vsol$, and a typical peculiar velocity for the high-mass star forming regions (HMSFRs) associated with the maser sources, as free parameters when fitting our models (though we place a prior on $\vsol$). This allows us to investigate claims of a lag behind circular rotation associated with the HMSFRs, and whether there is any evidence that the peculiar velocity of the Sun found from stars in the Solar neighbourhood \citep[by][]{SBD10} is inappropriate when looking at the larger scales of the Galaxy. We find no evidence of any significant lag associated with the maser sources. There is a strong degeneracy between the effect of a change to the assumed peculiar motion of the Sun and the assumed typical peculiar motion of the HMSFRs. We can therefore see that if $U_\odot\sim14\kms$ \citep[as suggested by][]{Sc12}, the HMSFRs are typically moving radially inwards with a velocity of ${\sim}6\kms$, and if $V_\odot\sim26\kms$ \citep{Boea12} then the typical HMSFR leads circular rotation by ${\sim}11\kms$. We argue that such significant deviations from circular velocities are unlikely for these objects.

Code required to compute the properties of our models and to compute orbits within them has been made available at  \url{https://github.com/PaulMcMillan-Astro/GalPot}, and in the appendix we give a brief description of its use.

With the further release of data from \Gaia, the number of stars for which we have accurate measurements of position and velocity is about to increase enormously. The data will enable us to improve our understanding of the Milky Way's structure dramatically, and improve upon these models. As a summary of our understanding of the Milky Way's potential in the pre-\Gaia\ era, the models provided by this study act as important starting point, and should be used to estimate the orbits of the stars observed by \Gaia, to help build our understanding of their place in the Galaxy.

\section*{Acknowledgements}
The author is grateful to former colleagues in the Oxford Dynamics group, and to Justin Read, for helpful conversations in the early stages of this study; to Peter Kalberla for clarification regarding the gas content of the Milky Way; to Stacy McGaugh for a helpful tweet; to Ortwin Gerhard for helpful comments; to Julien Lavalle for pointing out an error in unit conversion; and to Louise Howes for careful reading of the finished paper.  
Funding for the research in this study came 
from the European Research Council, the Swedish
National Space Board and the Royal Physiographic Society in Lund. The corner plots were created with the \textsc{corner.py} package \citep{corner}.  

\bibliographystyle{mnras} \bibliography{new_refs}

\appendix
\section*{Appendix: GalPot code and orbit integrations}
The \textsc{GalPot} code is designed to provide the gravitational potential associated with axisymmetric density distributions like those used in this study. It was originally created by Walter Dehnen for the study of DB98, which contains a description of how it determines the potential, and is available under a Gnu Public Licence. The version provided at \url{https://github.com/PaulMcMillan-Astro/GalPot} is there for convenience, and includes files that provide the parameters for the potentials of our main model (Table~\ref{tab:bestmodel}), models with all of the variations shown in Table~\ref{tab:altmodels}, the best fitting model of \cite{Piea14}, the `best' and `convenient' models from Paper I, and the four main models from DB98. We have included several example executables to demonstrate how this can be used, each of which provides helpful information about the required input if they are run without any arguments.

To use one of these potential in C++ code one needs to enter the line
\begin{verbatim}
#include "GalPot.h"
\end{verbatim}
in the declarations. Then, for example, to use our main model potential one could can use the code
\begin{verbatim}
ifstream from("pot/PJM16_best.Tpot");
Potential *Phi  = new GalaxyPotential(from);
from.close();
\end{verbatim}
The GalaxyPotential class can provide several useful properties of the potential 
(illustrated in the example executable {\tt testGalPot.cc}), with the main ones being the potential and its derivatives at a point $R,z$, so for example the following code fragment
\begin{verbatim}
double R=8.2, z=1.1;
double P, dPdR, dPdz;
P = (*Phi)(R,z);
// or if derivatives needed too...
P =  (*Phi)(R,z,&dPdR,&dPdz);
\end{verbatim}
provides the potential at $R,z$ (as {\tt P}) and its derivatives with respect to $R$ and $z$ at that point ({\tt dPdR} and {\tt dPdz} respectively). Note that it does not return the force per unit mass on an object at $R,z$ (which is of course {\tt -dPdR} and {\tt -dPdz} in the two directions).

All inputs and outputs of {\tt GalPot} are in the code units, which are radians, kpc, $\mathrm{Myr}$ and ${\rm M}_\odot$. Conversions are provided by the {\tt Units} namespace (found in the file {\tt Units.h}), and the format is that to put a value into code units you write {\tt ValueInCodeUnits = ValueInOriginalUnits * Units::NameOfUnit}, so, for example
\begin{verbatim}
double v = 230. * Units::kms;
\end{verbatim}
gives the v the value of $230\kms$ in code units (note that this is ${\sim}0.23$ because $1\kms\approx1\pc\Myr^{-1}$). Equivalently, of course, one can write, for example
\begin{verbatim}
std::cout <<  v / Units::kms;
\end{verbatim}
to convert back from code units, and output the value in $\kms$.

We also provide simple Runge-Kutta orbit integration routines to determine properties of orbits in a given potential. Example executable files demonstrate their use. For example, the executable {\tt findOrbitProperties.exe} finds the properties of an orbit in our best fitting model, given input initial $R,z$ (in kpc) and $v_R,v_z,v_\phi$ (in $\kms$). So, to find the properties of the Sun's orbit, taking $R_0$ and $v_0$ from Table~\ref{tab:mainresults}, $z_0$ from \cite{JJBGeSp97} and the Solar peculiar velocity from \cite{SBD10}, one would run
\begin{verbatim}
./findOrbitProperties.exe 8.2 0.014 11.1 7.25 -245
\end{verbatim}
and receive the output 
\begin{verbatim}
Guiding Centre radius: 8.62121
Minimum, Maximum Cylindrical radius: 8.11988,9.18502
Maximum z: 0.104202
Minimum, Maximum Spherical radius: 8.12024,9.18521
Energy: -153122 km^2/s^2
Angular Momentum: -2009 kpc km/s
Mean Cylindrical radius: 8.66851
\end{verbatim}
The executable {\tt findManyOrbitProperties.exe} does the same for a list of positions and velocities, in the same format, given in an input file, and outputs the results to a table.

The class {\tt OrbitIntegratorWithStats} is used by these executables to determine statistics about the orbits. For example:
\begin{verbatim}
Vector <double,6> XV;
XV[0] = atof(argv[1]) * Units::kpc; // R
XV[1] = atof(argv[2]) * Units::kpc; // z
XV[2] = 0.* Units::degree; // phi unimportant
XV[3] = atof(argv[3]) * Units::kms; // v_R
XV[4] = atof(argv[4]) * Units::kms; // v_z
XV[5] = atof(argv[5]) * Units::kms; // v_phi

// Set up orbit integrator, with integration time
// 10000 Myr
OrbitIntegratorWithStats OI(XV,Phi,10000.);

// Run integration
int fail = OI.run();

if(fail) {
 std::cout << "Orbit unbound"
} else {
 // output statistics, e.g.
 std::cout << "Maximum z: " << OI.Maxz << "kpc\n"
}
\end{verbatim}

\begin{figure}
  \centerline{\resizebox{0.7\hsize}{!}{\includegraphics{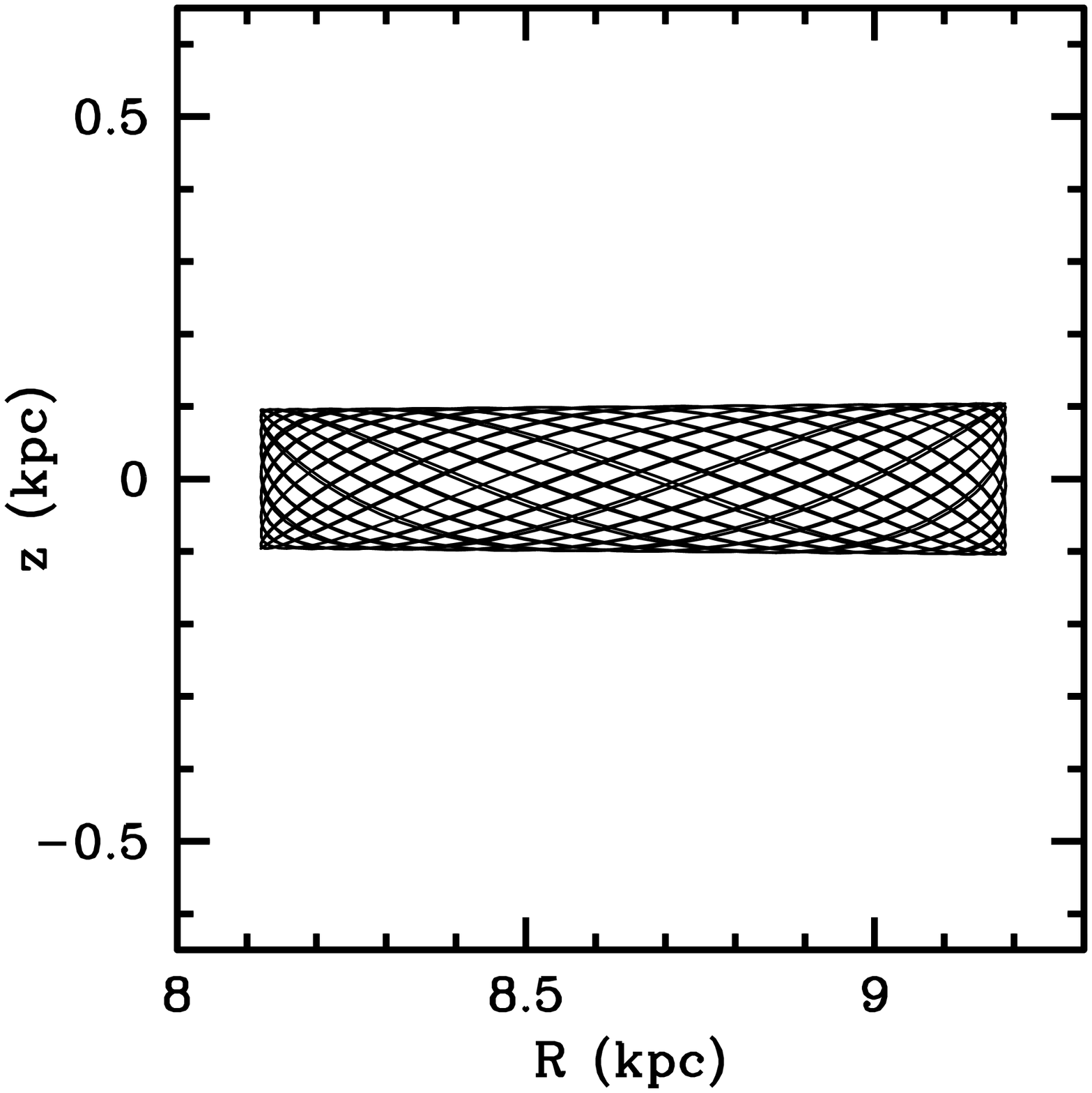}}}
  \caption{Trace of the Sun's orbit in the $R$-$z$ plane as determined in the 'best' potential found in this study, using the executable {\tt findOrbit.exe}. Only the first $1\,500$ points (representing $3\Gyr$) are shown, 
\label{fig:Rz}
}
\end{figure}

To output the track for an orbit integration (at points roughly evenly spaced in time), one can use another function from {\tt OrbitIntegratorWithStats} called {\tt runWithOutput}, which is demonstrated in the executable {\tt findOrbit.exe}. For example, to trace the orbit of the Sun, outputting $5000$ points over the full integration (which covers $10\Gyr$ by default), one would run
\begin{verbatim}
./findOrbit.exe 8.2 0.014 0 11.1 7.25 -245 \  
              SunOrbit.tab 5000
\end{verbatim}
The output table (in this case SunOrbit.tab) contains the full phase-space coordinates $(R,z,\phi,v_R,v_z,v_\phi)$. The path in the $R$-$z$ plane that this traces is plotted in Figure~\ref{fig:Rz}.

We also include code that can calculate coordinate transformations between commonly used coordinate systems, most notably galactic coordinates or equatorial coordinates. Conversions are done using the class {\tt OmniCoords}, and demonstrations of its use are provided in the example executables, such as {\tt findManyOrbitPropertiesfromEquatorial.exe} works in the same way as  {\tt findManyOrbitProperties.exe} but works in equatorial coordinates (by default, epoch J2000).

Note that by default {\tt OmniCoords} assumes that the Sun is at $R_0 = 8.21\kpc$, $z_0=0.014\kpc$, that the circular speed at $R_0$ is $233.1\kms$, that the solar peculiar motion is that determined by \cite{SBD10}, and that equatorial coordinates are J2000. These can be changed before converting any coordinates as follows:
\begin{verbatim}
OmniCoords OC;
OC.change_sol_pos(8.3,0.025);
OC.change_vc(-244.5*Units::kms);
OC.change_vsol(10*Units::kms,5.25*Units::kms,
                7.17*Units::kms);
OC.change_epoch(2015.0);                
\end{verbatim}
Note also that the direction that the Sun rotates around the Galactic centre is the \emph{negative} $\phi$-direction (to ensure that the Galactic north pole is the positive $z$-direction and the Sun is at $\phi=0$).

We also include code that can provide statistics on the orbits associated with observed stars, taking uncertainties into account. We provide an example executable called {\tt findManyOrbitPropertiesfromEquatorialwErrors.exe}, which takes input in equatorial coordinates, with uncertainties (assumed to be Gaussian), and provides the parameters of the orbits, with uncertainties. It achieves this by Monte Carlo sampling within the uncertainties (with the number of samples chosen by the user), and integrating the orbits with the resulting starting positions and velocities to determine the orbital parameters. The statistics that are output are found as the median value from the Monte Carlo, with separate uncertainties in the positive and negative directions found by taking the $15.87$ and $84.13$ percentile values (the per cent equivalent of 1$\sigma$). It is worth noting that we \emph{have} to take the median value, as the expectation value of many quantities (such as apocentric distance) is invariably infinite if the uncertainties are Gaussian. This is due to the fact that a finite change in any of the coordinates (which has a finite probability) can make the orbit unbound (and therefore having an infinite value of, for example, apocentric distance). We also include code that derive the same properties for input coordinates that come from \Gaia\ (i.e. parallax), or from the RAVE survey's multi-Gaussian fits to distance modulus pdfs.

The bulk of this code is also in the Torus Mapper package, and therefore described in section 2 of \cite{JJBPJM16}. We have separated it out and added the utilities for orbit integration for ease of use (the \textsc{GalPot} package is referred to as \textsc{falPot} in the Torus Mapper package for historical reasons). However, it remains the case that better understanding and characterisation of orbits comes from the use of action-angle coordinates, which is beyond the scope of this simple package of code -- see \cite{SaJJB16} for a review of methods available for calculating actions.

\bsp	
\label{lastpage}
\end{document}